# Thermodynamics of nonequilibrium processes in a tornado: synergistic approach

**G.P. Bystrai, I.A. Lykov, S.A. Okhotnikov**

Ural Federal University,Yekaterinburg, Russia
Gennadyi.Bystrai@usu.ru
[Submitted in journal "Modern physics" 20 July 2011]

**Abstract**

In the mathematical modeling of strongly nonequilibrium and nonlinear processes in a tornado approach based on the momentum transfer equations with a model function of sources and sinks is used, which puts this approach to the sharpening problems, where the maximum velocity distribution over the space of indefinitely growing for a limited time. Nonlinear momentum source in the medium leads to a blow-up regime, and the development of the regime, itself generated by a nonlinear medium, leads to self-organization, which is described by numerical methods. In this case, the competition between processes of the increasing due to nonlinear source and pulse propagation taking into account the viscosity of the medium leads to the appearance of new medium characteristic – some linear size – the spatial diameter of the tornado, on which these processes "balance" each other.

The approach allows to obtain equations that describe the physical effects observed and explain nonlinear transfer mechanisms of layering momentum in a tornado, determining the speeds and pressure gradients for nonpotential flow, as well as each altitude local layer thermodynamic properties – entropy production, entropy change rate and free energy. We describe the spatial and velocity characteristics of tornado various intensities, speed components in the tornado core and its surroundings, the emergence of spiral waves, rising to a arm height, the appearance of spatial ring (thrombus) and mesovortexes outside.

**Introduction**

Tornado (sp. Tornado «twister») – an atmospheric vortex, which arises in cumulonimbus (thunderstorm) cloud and spreads down to the ground, in the form cloud arms or trunk of tens or hundreds of meters. It is believed that the formation of a tornado form cloud distinguishes it from some of the seemingly similar and also of the various phenomena in nature, such as tornado-vortices and dust (sand) vortices [12]. Velocity drop is so high that the closed gas-filled objects, including buildings, explode from the inside because of the pressure difference. This phenomenon increases the damage from the tornado, obstructs to determine the parameters therein. Most estimates of this value are known from indirect observations. Depending on the intensity of the vortex flow velocity in it may vary. It is believed to exceed 18 *m*/*s* and can reach, according to some indirect estimates, 1,300 *km*/*h*. The tornado moves along with its generating cloud. This movement can provide speeds of tens of *km*/*h*, usually 20-60 *km*/*h*. It is estimated that the energy of an ordinary tornado radius of 1 *km* and an average speed of 70 *m/s* is comparable with the energy of standard nuclear bomb. Time Record of tornado existence can be considered as Mettunsky tornado that May 26, 1917 7 hours 20 minutes walked through the territory of the U.S.A. about 500 *km*, killing 110 people. The width of this tornado fuzzy funnel was 0.4-1 *km*, inside it whip-like funnel was visible. Another famous case of a tornado is a tornado three States (Tristate tornado), which is March 18, 1925 has passed through the states of Missouri, Illinois and Indiana, worked his way 350 *km* in 3.5 hours. Its fuzzy funnel diameter ranged from $\ell_0$ of 800 *m* to 1.6 *km*.

At the contact point of tornado funnel (arm) base with the ground or water can occur *cascade* – a cloud or a cloud of dust, debris and objects dredged from land-whether or splashing. In the formation of tornado observer sees descending from the sky funnel the cascade from the ground up, which then encompasses the bottom of the funnel. Funnel do not touching the ground, can envelop *holder*. Merging, cascade, holder, and the parent cloud creates the illusion of a larger funnel than it actually.

Although the numerical simulation of tornado in view of many approximations performed[2], yet the reasons for the tornado formation are not fully understood until now [9], since there is no complete mathematical models of blow (sharpening mode) that include thermodynamics nonequilibrium processes and would explain the tornado core occurrence. You can specify only some general information, the most characteristic for typical tornadoes. Tornadoes in its development which exactly corresponds to the blow-up regime are pass through several basic stages. At the initial stage from thunderstorm cloud the initial funnel appears, hanging above the ground. Cold air layers directly under the cloud rush down to replace the warm air, which in turn rise up. Such unstable system usually formed when two atmospheric fronts connected – warm and cold. The potential energy of this system goes into kinetic energy of rotational movement of air.

As "ingredients" for get tornado scientists call two of atmospheric phenomena. The first phenomenon is front air stratification for warm moist air near the ground and cold air in the upper atmosphere. The second component is the difference in the wind speed and direction at different altitudes. When warm moist air collides with colder air mass flow, which moves with different velocity and in the other direction, tornado arises. Nevertheless, we need information on full condition for occurrence of a tornado, and thermodynamic conditions of existence of the stationary regimes.

Tornado air masses usually rotate counter-clockwise with a velocity of 450 *km/h*, pulling into the vortex water, dust, all objects appearing on its path, and transporting them over long distances. The strongest tornadoes occur in the U.S.A [16,23].

It is believed that the famous lifting force of tornado is not due to the fact that the funnel sucks in itself subjects, and because of the fact that the rotating air column has a vertical component in the core – the vertical turbulence. This means that the air encircles funnel is not strictly in a circle (then the vertical movement would be missing) and not at a fixed spiral (then the vertical movement would be constant), and has rapidly changing vertical velocity component $\vartheta_z$. The theoretical description of the air velocity in the funnel and in general description of the whole of physics processes and, in particular, the thermodynamics of strongly non-equilibrium processes is still presents a serious problem. This is the issue the subject of this article.

### 1.Model assumptions

**1. Hypotheses about the vertical velocity component.** In the central part of the tornado a core width of 100-150 *m* or less presents, in which the powerful – up to 60-80 *m/s* – air downflows observed. Considerably-cooled descending air (Fig. 1) converge at the surface, thus increasing the destructive tornado force with the base formation. The huge - up to 70-80 *m* – updrafts around the core is observed. Thus the relative velocities can reach the value of ~ 200 - 300 *m/s*. Water vapor condensation takes place throughout of visible tornado length (arm), which may explain the whitish color appearance, which is visible from afar. When a tornado takes dust and sand – it gets dark.

Tornado is moving at high speed air masses, which are trying to fill his inner core, because of low pressure developed in areas with great speed, but they are so twisted around an axis in angular momentum conservation law, that centrifugal force don't allow them to the center.

We select a unit volume thin layer which is normal to the tornado arm. There are movement sources and sinks in this layer, which depend on the horizontal velocities $\vartheta_x(x,y,t), \vartheta_y(x,y,t)$ and the velocity modulus $\vartheta = \sqrt{\vartheta_x^2 + \vartheta_y^2}$. Problem will be solved under the assumption of the isothermal arm layer and the surrounding area $T_0 \cong const$. This substantially simplifies the problem in the original formulation. Further we will use the following hypotheses to solve the problem.

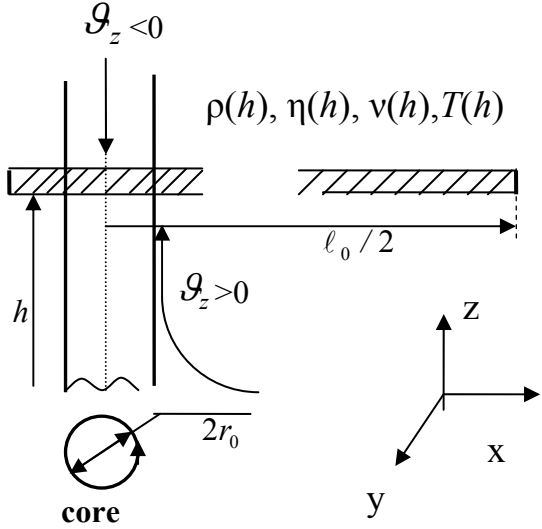

Fig. 1. The hypothesis about the vertical velocity component. The parameters $\rho(h)$ − the density, $\eta(h)$ − the dynamic and $\nu(h)$ − the kinematic viscosity; $T(h)$ − the temperature, which are the functions of altitude.

**2. The energy conservation law for a unit volume layer.** Let the $F(\theta(\xi_e,\xi_i),V,t)$ is Helmholtz free energy function for a planar nonequilibrium unit volume ($V$) air layer, which is the system state function and takes the minimum value $F_0$ in the equilibrium; here $\theta$ – nonequilibrium temperature value, $\xi_i$ – internal and $\xi_e$ – external variables. We shall call them the disequilibrium parameters. Then for the nonequilibrium state function – the layer free energy $F(\theta(\xi_e,\xi_i),V,t)$, which is assumed twice differentiable, – the total time derivative is [18]:

$$\frac{dF}{dt} = \frac{\partial F}{\partial \theta}\frac{d\theta}{dt} + \frac{\partial F}{\partial \theta}\frac{\partial \theta}{\partial \xi_e}\frac{d\xi_e}{dt} + \frac{\partial F}{\partial \theta}\left(\frac{\partial \theta}{\partial \xi_i}\right)\frac{d\vec{\xi_i}}{dt} + \frac{\partial F}{\partial V}\frac{dV}{dt} + \frac{\partial F}{\partial t}.$$

Arrows indicate vector quantities. Guided by the physical sense we introduce the following notation:

$$\frac{\partial F}{\partial \theta} = -S, \quad \frac{\partial F}{\partial V} = -P; \quad \vec{J_i} = -\frac{d\vec{\xi_i}}{dt}, \quad J_e = -\frac{d\xi_e}{dt},$$

$$\frac{\partial F}{\partial \xi_e} = \frac{\partial F}{\partial \theta}\frac{\partial \theta}{\partial \xi_e} \equiv -S\frac{\partial \theta}{\partial \xi_e}, \quad \frac{\partial F}{\partial \xi_i} = \frac{\partial F}{\partial \theta}\frac{\partial \theta}{\partial \xi_i} \equiv -S\frac{\partial \theta}{\partial \xi_i};$$

$$\vec{X_i} = -\frac{S}{\theta}\left(\frac{\partial \vec{\theta}}{\partial \xi_i}\right) = -\frac{1}{T}\vec{\nabla}\vartheta, \quad X_e = \frac{S}{\theta}\frac{\partial \theta}{\partial \xi_e}; \quad \sigma = -\frac{1}{\theta}\frac{\partial F}{\partial t}.$$

Here $S$ – nonequilibrium entropy, $P$ – thin flat air layer pressure, which determine the nonequilibrium state along with other parameters, $\theta\sigma = const$ – non-formalized energy loss, which is always takes place in open systems. For the momentum transfer problem internal variables take the form $\xi_i = \hat{\eta}\hat{\varepsilon}$ – the vector variable, $\hat{\varepsilon}$ – the strain tensor. $J_e$ and $J_i$ – external and internal thermodynamic flows. $X_e$ and $X_i$ – external and internal thermodynamic forces.

Then for the finite local layer the equation of the free energy change rate can be written as:

$$\frac{dF}{dt} = -S\frac{\partial \theta}{\partial t} - P\frac{dV}{dt} + \theta J_e X_e - \theta \vec{J_i}\vec{X_i} - \theta\sigma.$$

This is the mathematical expression of the local imbalance principle with non-formalized energy losses. This differential equation is applicable to both linear and nonlinear processes and expresses in the local form variation of the energy.

Separating nonequilibrium part $\delta T$ in the temperature $\theta = T \pm \delta T$, which is responsible for the cooling or heating, we reduce this equation to the form

$$\frac{d(F - F_0)}{dt} = -T\left(\sigma^e + \vec{J}_i \vec{X}_i + \sigma\right),$$

when $\dfrac{dF}{dt} \equiv \dfrac{dF}{dt} \pm \delta T(-X_e J_e + X_i J_i) \pm \delta T \sigma$ – nonequilibrium free energy in the layer, in the equilibrium $F = F_0$ and $\dfrac{dF_0}{dt} = -S\dfrac{\partial T}{\partial t} - P\dfrac{dV}{dt}$, $\sigma^e \equiv -J_e X_e$. Temperature – some medium uniform temperature of the layer, depending on the altitude, but not on the horizontal coordinates.

We start from the resulting free energy change rate equation for unit volume thin flat layer with movement sources and sinks (layer is normal to the tornado arm axis) which for a constant volume ($V$ = const) and the complete absence of energy loss ($\sigma \equiv 0$) at $T = T_0$ represents in scalar form:

$$\frac{dF}{dt} = -T_0\left(\sigma^e + \vec{J}_i \vec{X}_i\right), \quad \text{where} \quad \frac{dF}{dt} = \left(\frac{\partial \vec{F}}{\partial \vec{\vartheta}}\right)_V \frac{\partial \vec{\vartheta}}{\partial t}. \tag{1}$$

In equation (1) $T_0 \sigma^e \lessgtr 0$ – function of sources and sinks ($[\sigma^e] = J/K \cdot s \cdot m^3$), $\vec{J}_i \vec{X}_i$ – the scalar product of thermodynamic fluxes and forces (entropy of production), $T_0 \equiv T_h$ – a mean uniform layer temperature, which depends on altitude but don't depend on the coordinates $x$, $y$; $\vec{\vartheta}$ – the local velocity value of the continuum in the selected layer, $[\vartheta] = m/s$, $[\eta] = Pa \cdot s$, $[F] = J/m^3$. The expression in brackets characterizes the entropy change rate for the layer volume by Prigogine [7,8]:

$$\frac{dS}{dt} = \frac{d_e S}{dt} + \frac{d_i S}{dt} \equiv \sigma^e + \sigma^i, \quad \sigma^i = \vec{J}_i \vec{X}_i.$$

Equation (1) reflects the fact that the self-organization – the dissipative structures formation, in which the entropy decrease occurs, results to the layer free energy increases, and their destruction – the entropy increase occurs when the free energy decreases.

Recording in the form (1) implies the possibility of height integrating and determinat of the free energy for the entire volume occupied by a tornado.

**3. Thermodynamic fluxes and forces.** For the two-dimensional case considered the velocity vector has two components in the spatial coordinates $\vec{\vartheta} = (\vartheta_x, \vartheta_y)$, where $\vartheta_x(x,y,t), \vartheta_y(x,y,t)$. We introduce a viscosity in the second-rank tensor form

$$\hat{\eta} = \begin{pmatrix} \eta_1 & -\eta_2 \\ \eta_2 & \eta_1 \end{pmatrix},$$

where $\eta_1$ – the own kinematic viscosity, $\eta_2$ – viscosity associated with the symmetric interaction of the velocity components $\vartheta_x, \vartheta_y$ through the cross-coefficients, which characterize this interaction in the layer. Generated non-equilibrium hydrodynamic process outside the arm in the layer is characterized by the thermodynamic force $\vec{X}_i$ ($[X_i] = 1/s \cdot K$) and flow $\vec{J}_i$ ($[J_i] = N/m^2$), which are vectors and can be represented as:

$$\vec{X}_i = -\frac{1}{T_0}\vec{\nabla}\vartheta, \quad \vec{J}_i = L_{ik}\vec{X}_k = -\hat{\eta}\vec{\nabla}\vartheta. \tag{2}$$

Here $L_{ik} = \hat{\eta} T_0$ – Onsager coefficient ($[L_{ik}] = Pa \cdot s \cdot K$), associated with the tensor viscosity.

Then in the flat layer the linear approximation of the thermodynamic flows and forces preset in the form of expressions

$$J_x = -\eta_1 \frac{\partial \vartheta_y}{\partial x} + \eta_2 \frac{\partial \vartheta_x}{\partial x} \qquad X_x = -\frac{1}{T_0}\left(\frac{\partial \vartheta_y}{\partial x} - \frac{\partial \vartheta_x}{\partial x}\right)$$
$$J_y = -\eta_2 \frac{\partial \vartheta_y}{\partial y} - \eta_1 \frac{\partial \vartheta_x}{\partial y} \qquad X_y = -\frac{1}{T_0}\left(\frac{\partial \vartheta_y}{\partial y} + \frac{\partial \vartheta_x}{\partial y}\right) \qquad (3)$$

which characterize the feedbacks between them – the interaction (mutual influence) flux transport stream components $J_x, J_y$.

**4. Compatible with the Navier-Stokes equations.** After time differentiation of (1), and also taking into account of the momentum flux variation principle from [14], we obtain a differential equation for the free energy change second derivative

$$\frac{d^2 F}{dt^2} = -T_0\left(\frac{\partial \sigma^e}{\partial t} + \frac{\hat{\eta}}{T_0}\vec{\nabla}\vec{\vartheta}\vec{\dot{\vartheta}}\right).$$

In other side

$$\frac{d^2 F}{dt^2} = \frac{d}{dt}\left(\frac{\partial F}{\partial \vartheta}\right)_V \frac{d\vec{\vartheta}}{dt} + \left(\frac{\partial F}{\partial \vartheta}\right)_V \frac{d^2 \vec{\vartheta}}{dt^2}, \quad \text{where} \quad \left(\frac{\partial F}{\partial \vartheta}\right)_V = -\rho\vec{\vartheta}, \quad \frac{d\vec{\vartheta}}{dt} = \frac{\partial \vec{\vartheta}}{\partial t} + \left(\vec{\vartheta}\vec{\nabla}\right)\vec{\vartheta};$$

here $\rho$ – the air density in the layer ($[\rho] = kg/m^3$), which will depend on the tornado layer at th altitude $h$. From last two expressions we obtain the hyperbolic equation

$$\rho\frac{d\vec{\vartheta}}{dt}\frac{d\vec{\vartheta}}{dt} + \rho\vec{\vartheta}\frac{d^2\vec{\vartheta}}{dt^2} = \hat{\eta}\vec{\nabla}\vec{\vartheta}\cdot\frac{d\vec{\vartheta}}{dt} + T_0\frac{d\sigma^e}{dt}, \qquad (4)$$

If the principle of spatial locality is performed, i.e. $\frac{1}{\Delta\tau}\left(\frac{\partial F}{\partial \vartheta}\right) \ll \frac{d}{dt}\left(\frac{\partial F}{\partial \vartheta}\right)$ or $\frac{\vec{\vartheta}}{\Delta\tau} \ll \frac{d\vec{\vartheta}}{dt}$, where $\tau$ – the stress relaxation time, then the left side second term can be neglect in the original formulation of the problem and its solution. As a result from the last equation we obtain the limiting parabolic equation for momentum transport in the form of Euler

$$\frac{\partial \vec{\vartheta}}{\partial t} = \hat{\nu}\Delta\vec{\vartheta} + \frac{T_0}{\rho}\left(\frac{\partial \sigma^e}{\partial \vartheta}\right). \qquad (5)$$

Saving the second term leads to the hyperbolic equation of the momentum transfer:

$$\frac{\partial \vec{\vartheta}}{\partial t} + \tau\frac{\partial^2 \vec{\vartheta}}{\partial t^2} = \hat{\nu}\Delta\vec{\vartheta} + \frac{T_0}{\rho}\left(\frac{\partial \sigma^e}{\partial \vartheta}\right).$$

Equation (5) is compatible with the Navier-Stokes equations in vector form

$$\frac{\partial \vec{\vartheta}}{\partial t} = \hat{\nu}\Delta\vec{\vartheta} + \frac{1}{\rho}\left(\xi + \frac{1}{3}\eta\right)graddiv\vec{\vartheta} - \frac{1}{\rho}\vec{\nabla}p + \vec{f} - \left(\vec{\vartheta}\vec{\nabla}\right)\vec{\vartheta}$$

for an incompressible fluid ($div\vartheta = 0$) and fulfillment of the following equation for the function of movement sources $\sigma^e$ in the vector form equation

$$\frac{T_0}{\rho}\left(\frac{\partial \vec{\sigma}^e}{\partial \vartheta}\right) = -\frac{1}{\rho}\vec{\nabla}p + \vec{f} - \left(\vec{\vartheta}\vec{\nabla}\right)\vec{\vartheta}, \qquad (6)$$

where $p$ – pressure, $\vec{f}$ – the volume forces ($[f] = m/s$). In these expressions $\hat{\nu} = \hat{\eta}/\rho$ – the tensor of the kinematic viscosity ($[\nu] = m^2/s$). Characteristics of models such as the kinematic viscosity, air density and pressure are the functions of the altitude $h$ according to [11]

$$p(h) = p_c\left(1 - \frac{Lh}{T_0}\right)^{\frac{g\mu}{RL}}, \quad \rho(h) = \frac{p(h)\mu}{RT_0}$$

where $R$ − universal gas constant ($[R] = J/K\,mol$), $k$ – Boltzmann constant ($[k] = J/K$), $\mu$ − molar mass of air ($\mu = 29 \cdot 10^{-3}\,kg/mol$), $\rho_c$, $p_c$ – density and pressure of the sea-level at $h = 0$, $L$ − the temperature fall rate with altitude ($L = 0.0065\,K/m$). Kinematic viscosity, in contrast to the density and pressure, increases with altitude increasing, what is important to consider layered problem solution.

**Entropy production.** On the basis of (3) the entropy production expression for the consideration an infinitely thin layer in this case takes the form:

$$\sigma^i(h) = J_x X_x + J_y X_y = \frac{\eta_1(h)}{T_0}\left[\left(\frac{\partial \vartheta_y}{\partial x}\right)^2 + \left(\frac{\partial \vartheta_x}{\partial y}\right)^2\right] +$$

$$+ \frac{\eta_2(h)}{T_0}\left[\left(\frac{\partial \vartheta_x}{\partial x}\right)^2 + \left(\frac{\partial \vartheta_y}{\partial y}\right)^2\right] - \frac{\eta_1(h) + \eta_2(h)}{T_0}\{\vartheta_x, \vartheta_y\}, \quad (7)$$

where

$$\{\vartheta_x, \vartheta_y\} = \frac{\partial \vartheta_x}{\partial x}\frac{\partial \vartheta_y}{\partial x} - \frac{\partial \vartheta_x}{\partial y}\frac{\partial \vartheta_y}{\partial y}.$$

Entropy production ($[\sigma^i] = J/K \cdot s \cdot m^3$) is the positive function (Lyapunov function), since $\sigma^i \geq 0$, $\dot\sigma^i \leq 0$. Positivity condition of the quadratic form (7) corresponds to the inequality imposed on the viscosity coefficients: $\eta_1^2 \geq \eta_2^2$ [7].

**5. The hypothesis of the movement sources and sinks presence in the layer as a statement of the sharpening problem.** We will solve this problem as the problem with the so-called "sharpening". In this formulation the formation of structures localized in space is possible, in which describable variable can indefinitely (or limited) increase. Following the ideas of Samarskiy with his students [2,4,5,20] in problems with sharpening the second term in the differential equation of motion (5) is presented as a sum of linear sources and nonlinear sinks:

$$\frac{T_0}{\rho(h)}\left(\frac{\partial \vec{\sigma}^e(h)}{\partial \vec{\vartheta}}\right) = q\vec{\vartheta} - \hat{\alpha}|\vartheta|^2\vec{\vartheta}, \qquad \hat{\alpha} = \begin{pmatrix} \alpha_1 & -\alpha_2 \\ \alpha_2 & \alpha_1 \end{pmatrix}. \quad (8)$$

In this expression $q$ − constant characterizing the sources ($[q]=1/s$), $\hat{\alpha}$ – tensor describing the nonlinear sinks ($[\hat{\alpha}] = s/m^2$). Tensor nature of the nonlinear sink coefficient associated with the interaction between two horizontal velocity components, then the body forces are

$$\vec{f} = \frac{1}{\rho(h)}\vec{\nabla}p + \left(\vec{\vartheta}\vec{\nabla}\right)\vec{\vartheta} + q\vec{\vartheta} - \hat{\alpha}|\vartheta|^2\vec{\vartheta}. \quad (9)$$

All body forces include the pressure gradient and correspond to velocity sources and sinks functions.

**6. Reduction to dimensionless form.** We introduce the normalization parameters in the derived model: $x^* = x/\ell_0$, $y^* = y/\ell_0$, $t^* \equiv t/t_0$, $\vartheta^* = \vartheta/\vartheta_c$, where $\ell_0, t_0, \vartheta_c$ − the normalization parameters (scales) of the coordinates, time and velocity, respectively. According to the notation the limiting parabolic equation for momentum transfer in the form of Euler (5) can be reduced to dimensionless form

$$\frac{\partial \vec{\vartheta}^*}{\partial t} = \nu^*(h^*)\Delta \vec{\vartheta}^* + \frac{1}{\rho^*(h^*)}\left(\frac{\partial \vec{\sigma}^{e*}(h^*)}{\partial \vartheta^*}\right),$$

where $\nu_1^*(h) = \dfrac{\nu_1(h)t_0}{\ell_0^2}$, $\nu_2^*(h) = \dfrac{\nu_2(h)t_0}{\ell_0^2}$, $\sigma^{e*}(h^*) = \sigma^e(h)\dfrac{T_0 t_0}{\rho_c \vartheta_c^2}$, $\rho^*(h^*) = \dfrac{\rho(h)}{\rho_c}$, $h^* = \dfrac{h}{h_0} = \dfrac{\mu g h}{RT_0}$.

**7. The pressure gradient for nonpotential flows in tornado.** From expressions (8) and (9) it is follow that the pressure gradient is associated with source function:

$$-\frac{b}{\rho^*(h^*)}\vec{\nabla}p^* = -\vec{f}^* + \left(\vec{\vartheta}^*\vec{\nabla}\right)\vec{\vartheta}^* + \frac{1}{\rho^*(h)}\left(\frac{\partial \vec{\sigma}^{e*}}{\partial \vartheta^*}\right). \qquad (10)$$

Here the source function is

$$\frac{1}{\rho^*(h^*)}\left(\frac{\partial \vec{\sigma}^{e*}(h^*)}{\partial \vartheta^*}\right) = q^*\vec{\vartheta}^* - \hat{\alpha}^*|\vartheta^*|^2 \vec{\vartheta}^*,$$

where $\quad q^* = qt_0, \quad \hat{\alpha}^* = \hat{\alpha}t_0\vartheta_c^2, \quad p^*(h^*) = \dfrac{p(h)}{p_c}, \quad b = \dfrac{t_0 p_c}{\rho_c \vartheta_c \ell_0}, \quad \vec{f}^* = \vec{f}\dfrac{t_0}{\vartheta_c}.$

From (10) it results that for nonpotential flow the expression for pressure gradient in the flat air layer is:

$$-\frac{b}{\rho^*(h^*)}\vec{\nabla}p^* = -\vec{f}^* + \left(\vec{\vartheta}^*\vec{\nabla}\right)\vec{\vartheta}^* + q^*\vec{\vartheta}^* - \hat{\alpha}^*|\vartheta^*|^2 \vec{\vartheta}^*. \qquad (11)$$

Consequently, the function of the pressure gradient $\vec{\nabla}p^*(x^*, y^*)$ in a flat layer $h$ for nonpotential flow can be restored from a known velocity $\vec{\vartheta}^*(x^*, y^*)$ in this layer. From the known function $\vec{\nabla}p^*(x^*, y^*)$ we may also restore $p^*(x^*, y^*)$ function. Equation (11) contains terms associated with the source ($q^*\vec{\vartheta}^*$) and sinks ($-\hat{\alpha}^*|\vartheta^*|^2 \vec{\vartheta}^*$). Consider the body forces function in the form $\vec{f}^* = -\vec{\nabla}E_{nom} = -\vec{g}$, where $E_{nom} = gh^*$ – the potential energy per unit volume. We use the relation

$$\vec{\nabla}\left(\frac{\vartheta^{*2}}{2}\right) = \left(\vec{\vartheta}^*\vec{\nabla}\right)\vec{\vartheta}^* + \left[\vec{\vartheta}^*, rot\,\vec{\vartheta}^*\right],$$

then the equation (11) is converted to a form that speaks of the movement nonpotentiality:

$$-\frac{b}{\rho^*(h^*)}\vec{\nabla}p^* = \vec{\nabla}E_{nom} + \vec{\nabla}\left(\frac{\vartheta^{*2}}{2}\right) - \left[\vec{\vartheta}\times rot\,\vec{\vartheta}\right] + q^*\vec{\vartheta}^* - \hat{\alpha}^*|\vartheta^*|^2 \vec{\vartheta}^*, \qquad (12)$$

or $\qquad \left[\vec{\vartheta}\times rot\,\vec{\vartheta}\right] = \vec{\nabla}\left(E_{nom} + \dfrac{\vartheta^{*2}}{2} + \dfrac{b}{\rho^*(h^*)}p^*\right) + q^*\vec{\vartheta}^* - \hat{\alpha}^*|\vartheta^*|^2 \vec{\vartheta}^*.$

*Potential flow.* Consider the particular case of the potential flow, or flow that closed to the potential, when $rot\,\vec{\vartheta}^* = 0$. This means the validity of following relations

$$\frac{\partial \vartheta_x^*}{\partial y^*} = \frac{\partial \vartheta_y^*}{\partial x^*}, \quad \frac{\partial \vartheta^*}{\partial y^*} = \frac{\partial \vartheta_y^*}{\partial z^*}, \quad \frac{\partial \vartheta^*}{\partial x^*} = \frac{\partial \vartheta_x^*}{\partial z^*}.$$

If there are no sources and sinks, i.e. $q^* = 0$, $\hat{\alpha}^* = 0$, or they are modulo equal $q^*\vec{\vartheta}^* = \hat{\alpha}^*|\vartheta^*|^2 \vec{\vartheta}^*$, then we obtain the expression for the pressure gradient of the potential flow

$$-\frac{b}{\rho^*(h^*)}\vec{\nabla}p^* = \vec{\nabla}\left(\frac{\vartheta^{*2}}{2}\right) + \vec{\nabla}E_{nom}, \quad \text{or} \quad \vec{\nabla}\left(\frac{\vartheta^{*2}}{2} + \frac{b\cdot p^*}{\rho^*} + gh^*\right) = 0,$$

which corresponds to the Bernoulli equation in reduced form:

$$\frac{\vartheta^{*2}}{2} + \frac{b\cdot p^*}{\rho^*} + g^*h^* = const, \qquad (13)$$

where $p^*$ – dimensionless pressure in the area which wind speed is equal to $\vartheta^*$ on the altitude $h^*$, $g^* = gt_0/\vartheta_c$ – dimensionless gravitational acceleration ($m/s^2$). Constant in equation (13) is valid for an air layer at the altitude $h^*$.

**8. External entropy flows (source function).** We compute sources and sinks function further called outside sources function $\sigma^e(h)$. For this calculation we write the projections of equation (8) on the coordinate axes:

$$\begin{cases} \dfrac{T_0}{\rho(h)}\left(\dfrac{\partial \sigma^e(h)}{\partial \vartheta_x}\right) = q\vartheta_x - |\vartheta|^2(\alpha_1\vartheta_x - \alpha_2\vartheta_y) \\ \dfrac{T_0}{\rho(h)}\left(\dfrac{\partial \sigma^e(h)}{\partial \vartheta_y}\right) = q\vartheta_y - |\vartheta|^2(\alpha_2\vartheta_x + \alpha_1\vartheta_y) \end{cases}.$$

After integrating the first equation and finding the integration constant from the second for the flat layer at a certain altitude the absolute value of the external sources function becomes

$$\sigma^e(h) = \dfrac{d_e S}{dt} = \dfrac{q\rho(h)}{2T_0}\vartheta^2 - \dfrac{\alpha_1\rho(h)}{4T_0}\vartheta^4 - \dfrac{\alpha_2\rho(h)}{3T_0}\vartheta_x\vartheta_y\vartheta^2 ,$$

here $\vartheta = \sqrt{\vartheta_x^2 + \vartheta_y^2}$ – the horizontal direction velocity magnitude, $\rho(h)$ – parametric density function, depending on the altitude. According to the previous notation this function of external sources is reduced to dimensionless form:

$$\sigma^{e*}(h^*) = \dfrac{q^*\rho^*(h^*)}{2}\vartheta^{*2} - \dfrac{\alpha_1^*\rho^*(h^*)}{4}\vartheta^{*4} - \dfrac{\alpha_2^*\rho^*(h^*)}{3}\vartheta_x^*\vartheta_y^*\vartheta^{*2}. \qquad (14)$$

**2. Two-dimensional equations of motion in a plane layer.
Kuramoto-Tsuzuki equations for a tornado flat layer**

Used approach allows us to consider a more general case, including the interaction between velocity vector components through viscosity and sinks function in each layer. We write in dimensionless form the interdependent system of equations for the velocity components in case of incompressible fluid:

$$\begin{cases} \dfrac{\partial \vartheta_x^*}{\partial t} = v_1^*(h^*)\left(\dfrac{\partial^2 \vartheta_x^*}{\partial x^{*2}} + \dfrac{\partial^2 \vartheta_x^*}{\partial y^{*2}}\right) - v_2^*(h^*)\left(\dfrac{\partial^2 \vartheta_y^*}{\partial x^{*2}} + \dfrac{\partial^2 \vartheta_y^*}{\partial y^{*2}}\right) + q^*\vartheta_x^* - \left(\alpha_1^*|\vartheta^*|^2\vartheta_x^* - \alpha_2^*|\vartheta^*|^2\vartheta_y^*\right) \\ \dfrac{\partial \vartheta_y^*}{\partial t} = v_1^*(h^*)\left(\dfrac{\partial^2 \vartheta_y^*}{\partial x^{*2}} + \dfrac{\partial^2 \vartheta_y^*}{\partial y^{*2}}\right) + v_2^*(h^*)\left(\dfrac{\partial^2 \vartheta_x^*}{\partial x^{*2}} + \dfrac{\partial^2 \vartheta_x^*}{\partial y^{*2}}\right) + q^*\vartheta_y^* - \left(\alpha_1^*|\vartheta^*|^2\vartheta_y^* + \alpha_2^*|\vartheta^*|^2\vartheta_x^*\right) \end{cases}, \qquad (15)$$

where $x^* = x/\ell_0$, $y^* = y/\ell_0$ – dimensionless coordinates. Nonlinear terms depend on the parameters $v_1^*, v_2^*, q, \alpha_1^*, \alpha_2^*$ and describe the complex behavior of the horizontal velocity components $\vartheta_x^*, \vartheta_y^*$ for any selected air layer at the altitude $h$.

If we follow A. Turing [1], then we should expect that by solving a system of nonlinear equations (15) under certain conditions the spontaneous emergence of order in the form of dissipative structures can be detected. And if we follow the logic of the sharpening problems [4,5], we can claim that the velocities can indefinitely increase in a finite time and under certain conditions localized structures are arise in the environment.

**Solution of tornado problem within the bounds of sharpening problems.** We multiply the second equation in (11) by $i$ and add the left and right sides of the first and the second equations. The result is the following equation:

$$\dfrac{\partial \Phi}{\partial t} = v_1^*(1 + ic_1)\left(\dfrac{\partial^2 \Phi}{\partial x^{*2}} + \dfrac{\partial^2 \Phi}{\partial y^{*2}}\right) + q^*\Phi - \alpha_1^*(1 + ic_2)|\Phi|^2\Phi , \qquad (16)$$

where $\Phi = \vartheta_x^* + i\vartheta_y^*$ – velocity complex function, the constant $c_1 = v_2^*/v_1^*$ is related to viscosity and $c_2 = \alpha_2^*/\alpha_1^*$ is related to the sinks in the function of external sources. Equation (16) is parabolic type complex equation and is known as the Kuramoto-Tsuzuki equation [6]. This equation describes the regime with sharpening [3], so-called regime in which one or several variables characterizing the system indefinitely increase in a finite time (this time is called "sharpening time"). They are characterized by the appearance of spatially localized structures which are stable at constant external influences. Sharpening regimes give a approximate description (asymptotic) of many nonlinear systems types with strong positive feedback. They are typical to the problems in the theory of combustion and explosion, for some instabilities in plasma physics, for a number of processes studied by mathematical biophysics, hydrodynamics, chemical kinetics [21]. For the local layer at altitude $h$ with taking into account the function of sources and sinks the momentum conservation law holds [7].

The simplest solution of equation (16) provided by $q^* = \alpha_1^*$ is a homogeneous solution $\Phi = \alpha_2^* \exp(-ic_2 t^*)$. But this solution becomes unstable at a certain parameter $c_2$ decreasing, i.e. numerical solution is no longer asymptotically tends to the decision. With the loss of stability the symmetric solution $\Phi(y^*,x^*,t^*) = \Phi(x^*,y^*,t^*)$ and one-dimensional solution $\Phi(x^*,y^*,t^*) = \Phi(x^*,t^*)$ arise. This means that in some cases the behavior of the system in two-dimensional space can be determined by one-dimensional equation.

There is also a two-dimensional analogue of the self-similar solution $\Phi(x^*,y^*,t^*) = \Phi(x^*,t^*)$ of equation (16):
$$\Phi(x^*,y^*,t^*) = R(x^*,y^*)\exp(i\omega t^* + ia(x^*,y^*)).$$
The solution of this type is periodic, i.e. $\Phi(x^*,y^*,t^*) = \Phi(y^*,x^*,t^*+2\pi/\omega)$, and it is also self-similar. For such case the time dependence of the functions $R(\vartheta_x^{*2}(x^*,y^*,t^*) + \vartheta_y^{*2}(x^*,y^*,t^*))^{1/2}$ is omitted. Coefficients $r_{mn} = (a_{mn}^2 + b_{mn}^2)^{1/2}$ harmonic of the expansion
$$\Phi(x^*,y^*,t^*) = \sum_{m,n}^{\infty} (a_{mn} + ib_{mn})\cos\left(\frac{\pi n x^*}{\ell_0}\right)\cos\left(\frac{\pi n y^*}{\ell_0}\right)$$
time dependence is also absent.

**Spiral waves solution.** In two dimensional case equation (16) also has solutions, describing the so-called spiral waves [6]. At the transformation to polar coordinates $x = r\cos\varphi$, $y = r\sin\varphi$, solution of the equation (16) is a spiral wave, if $R(x^*,y^*)=R(r)$ and $a(x^*,y^*)=S(r)+m\varphi$, where $m=\pm1, \pm2, ...$ – represents the topological charge value. The case of $|m|>1$ describes a multiturn spiral wave. Zero value of topological charge $m = 0$ corresponds to a circular waves. The value of the topological charge determines the amount of revolving spiral waves "segments". But we will try avoid the term "arm" with respect to them in the future. The value of $m>0$ corresponds to the right twist waves, $m<0$ – left. It can be shown that changing the velocity values $\vartheta_x^*$ and $\vartheta_y^*$ at sign opposite is conditioned by the change of the topological charge sign. Following Bernoulli's equation for the velocity we can find other movement characteristics such as pressure and pressure gradients.

## 3. Thermodynamics of the transfer momentum processes with sources and sinks

Using the obtained expressions we can find the free energy change rate and entropy change rate, which allow us to identify the birth, development and attenuation of self-organization in a tornado on the thermodynamics-level in the local form (at the altitude). The thermodynamic approach for each layer at the altitude $h$ also allows us to find expressions for entropy production and entropy boundary flow across, as well as the free energy change rate.

**1. Entropy production** for the system (15) will be determined by expression (7), which can be written in dimensionless form

$$\sigma^{i*}(h^*) = \nu_1^*(h^*)\rho^*(h^*)\left[\left(\frac{\partial \vartheta_y^*}{\partial x^*}\right)^2 + \left(\frac{\partial \vartheta_x^*}{\partial y^*}\right)^2\right] +$$

$$+ \nu_2^*(h^*)\rho^*(h^*)\left[\left(\frac{\partial \vartheta_x^*}{\partial x^*}\right)^2 + \left(\frac{\partial \vartheta_y^*}{\partial y^*}\right)^2\right] - (\nu_1^*(h^*) + \nu_2^*(h^*))\rho^*(h^*)\{\vartheta_x^*, \vartheta_y^*\}. \quad (17)$$

Now we should turn to the Prigogine's theorem of the minimal entropy production [7]. At constant sources and sinks power and under constant pressure gradient and body forces ($\sigma^{e*} = const$) entropy production tends to decrease, taking the minimum positive value in the steady state in accordance with the dimensionless (reduced) equation which derived from (1) after time differentiation:

$$\frac{d\sigma^{i*}}{dt} = -\frac{d\sigma^{e*}}{dt} - \frac{d^2 F^*}{dt^2}, \quad \frac{\partial^2 F^*}{\partial t^2} \geq 0, \quad \frac{d\sigma^{i*}}{dt} \leq 0, \quad (18)$$

where dimensionless variables $\sigma^{i*} = \sigma^i \frac{T_0 t_0}{\rho_c \vartheta_c^2}$, $\sigma^{e*} = \sigma^e \frac{T_0 t_0}{\rho_c \vartheta_c^2}$, $F^* = F \frac{1}{\rho_c \vartheta_c^2}$, $t \equiv t/t_0$.

**2. External entropy flows** (a function of movement sources) is given by

$$\sigma^{e*}(h^*) = \frac{q^* \rho^*(h^*)}{2}\vartheta^{*2} - \frac{\alpha_1^* \rho^*(h^*)}{4}\vartheta^{*4} - \frac{\alpha_2^* \rho^*(h^*)}{3}\vartheta_x^* \vartheta_y^* \vartheta^{*2}. \quad (19)$$

**3. The entropy change rate** is

$$\frac{dS(h)}{dt} = \frac{d_e S}{dt} + \frac{d_i S}{dt} = \sigma^e(h) + \sigma^i(h) = \frac{q\rho(h)}{2T_0}\vartheta^2 - \frac{\alpha_1 \rho(h)}{4T_0}\vartheta^4 - \frac{\alpha_2 \rho(h)}{3T_0}\vartheta_x \vartheta_y \vartheta^2 +$$

$$+ \frac{\eta_1(h)}{T_0}\left[\left(\frac{\partial \vartheta_y}{\partial x}\right)^2 + \left(\frac{\partial \vartheta_x}{\partial y}\right)^2\right] + \frac{\eta_2(h)}{T_0}\left[\left(\frac{\partial \vartheta_x}{\partial x}\right)^2 + \left(\frac{\partial \vartheta_y}{\partial y}\right)^2\right] - \frac{\eta_1(h) + \eta_2(h)}{T_0}\{\vartheta_x, \vartheta_y\} \quad (20)$$

Parameters depend on the altitude $h$: $\rho(h)$, $\eta(h)$, $\nu(h)$. In dimensionless form the entropy change rate $\dot{S}^*(h)$ is equal to

$$\frac{dS^*(h^*)}{dt} = \sigma^{e*}(h^*) + \sigma^{i*}(h^*) = \frac{q^* \rho^*(h^*)}{2}\vartheta^{*2} - \frac{\alpha_1^* \rho^*(h^*)}{4}\vartheta^{*4} - \frac{\alpha_2^* \rho^*(h^*)}{3}\vartheta_x^* \vartheta_y^* \vartheta^{*2} +$$

$$+ \nu_1^*(h^*)\rho^*(h^*)\left[\left(\frac{\partial \vartheta_y^*}{\partial x^*}\right)^2 + \left(\frac{\partial \vartheta_x^*}{\partial y^*}\right)^2\right] + \nu_2^*(h^*)\rho^*(h^*)\left[\left(\frac{\partial \vartheta_x^*}{\partial x^*}\right)^2 + \left(\frac{\partial \vartheta_y^*}{\partial y^*}\right)^2\right] - (\nu_1^*(h^*) + \nu_2^*(h^*))\rho^*(h^*)\{\vartheta_x^*, \vartheta_y^*\}$$

where $\quad S^* = S \frac{T_0}{\rho_c \vartheta_c^2}$ ([$S$]=J/K·m$^3$). $\quad (21)$

**4. Self-organization conditions.** For the nonlinear system (plane layer) considered the following two main criteria of self-organization arising from (21) can be written: 1) $\sigma^{e*} < 0$, 2) $|\sigma^{e*}| \geq \sigma^{i*}$. This corresponds to an entropy decreasing in the layer $dS^*/dt < 0$, indicating the order growth and the emergence of dissipative structures – as a vortices system. As a result we obtain from (19) inequality for velocities $\vartheta_x^*, \vartheta_y^*$ symmetric function $f(\vartheta_x^*, \vartheta_y^*)$:

$$f(\vartheta_x^*, \vartheta_y^*) = \frac{\alpha_1^*}{4}\vartheta_x^{*2} + \frac{\alpha_2^*}{3}\vartheta_x^* \vartheta_y^* + \frac{\alpha_1^*}{4}\vartheta_y^{*2} - \frac{q^*}{2} > 0.$$

The symmetry of the function $f(\vartheta_x^*, \vartheta_y^*)$ indicates that the self-organization conditions is valid for $\vartheta_y^*$ and $\vartheta_x^*$. We find the discriminant of the equation $f(\vartheta_x^*) = 0$:

$$D = \left(\frac{\alpha_2^* \vartheta_y^*}{3}\right)^2 - \alpha_1^*\left(\frac{\alpha_1^*}{4}\vartheta_y^{*2} - \frac{q^*}{2}\right) = \frac{\alpha_1^{*2}}{4}\left(\left(\frac{4}{9}c_2^2 - 1\right)\vartheta_y^{*2} + \frac{2q^*}{\alpha_1^*}\right), \text{ where } c_2 = \frac{\alpha_2^*}{\alpha_1^*}.$$

Condition $f(\vartheta_x^*) > 0$ corresponds to a negative discriminant value $D < 0$, it is hold at

a) $-\frac{3}{2} < c_2 = \frac{\alpha_2^*}{\alpha_1^*} < \frac{3}{2}$;  
b) $|\vartheta_x^*| \geq \sqrt{\frac{2q^*}{\alpha_1^*(1 - 4c_2^2/9)}}$  
$|\vartheta_y^*| \geq \sqrt{\frac{2q^*}{\alpha_1^*(1 - 4c_2^2/9)}}$.

If we turn to the velocity magnitude, the inequality stands in the capacity of the self-organization condition:

$$\vartheta^{*2} \geq \frac{4q^*}{\alpha_1^*(1 - 4c_2^2/9)}, \quad c_2^2 = \frac{\alpha_2^{*2}}{\alpha_1^{*2}} < \frac{9}{4}. \tag{22}$$

This is the self-organization condition of the Kuramoto-Tsuzuki equation and this is the resulting boundary condition on the modulus of horizontal velocity in the layer, bearing in mind the basic condition of self-organization - an entropy decreasing in this volume. For example, when $q^*=1$, $\alpha_1^*=10$, $\alpha_2^*=1$ constant will be $c_2^2 = \frac{\alpha_2^{*2}}{\alpha_1^{*2}} = \frac{1}{100}$ and velocity magnitude – $\vartheta_{min}^* \geq 0.64$, i.e. magnitude must be equal to or greater than 0.64.

There is one self-organization condition limiting growth of the tornado transverse size $\ell_0$ – this condition is $\dot{S} = 0$ on the boundary, and it doesn't allow the transverse dimension to indefinitely grow. At this border competition between structures ordering and collapse leads to their alignment, which ultimately determines the existence of the tornado limited size, above which the self-organization is not observed.

**5. The free energy change rate. Integration law of the free energy conservation.** The free energy change rate $\dot{F}(h)$ by (1), (7), (14) is given by

$$\frac{dF(h)}{dt} = -T_0\left(\sigma^e(h) + \sigma^i(h)\right) = -T_0\left(\frac{q\rho(h)}{2T_0}\vartheta^2 - \frac{\alpha_1\rho(h)}{4T_0}\vartheta^4 - \frac{\alpha_2\rho(h)}{3T_0}\vartheta_x\vartheta_y\vartheta^2 + \right.$$
$$\left. + \frac{\eta_1(h)}{T_0}\left(\left(\frac{\partial\vartheta_y}{\partial x}\right)^2 + \left(\frac{\partial\vartheta_x}{\partial y}\right)^2\right) + \frac{\eta_2(h)}{T_0}\left(\left(\frac{\partial\vartheta_x}{\partial x}\right)^2 + \left(\frac{\partial\vartheta_y}{\partial y}\right)^2\right) - \frac{\eta_1(h) + \eta_2(h)}{T_0}\{\vartheta_x, \vartheta_y\}\right)$$

here we take into account that $T_0\sigma^e$ – the function of sources (if $\sigma^e > 0$) or sinks (if $\sigma^e < 0$). According to the previous notation after the transformation to dimensionless form we obtain dimensionless expression for the free energy change rate

$$\frac{dF^*(h^*)}{dt} = -\left(\sigma^{e*}(h^*) + \sigma^{i*}(h^*)\right) = -\left(\frac{q^*\rho^*(h^*)}{2}\vartheta^{*2} - \frac{\alpha_1^*\rho^*(h^*)}{4}\vartheta^{*4} - \frac{\alpha_2^*\rho^*(h^*)}{3}\vartheta_x^*\vartheta_y^*\vartheta^{*2} + \right.$$
$$\left. \nu_1^*(h^*)\rho^*(h^*)\left(\left(\frac{\partial\vartheta_y^*}{\partial x^*}\right)^2 + \left(\frac{\partial\vartheta_x^*}{\partial y^*}\right)^2\right) + \nu_2^*(h^*)\rho^*(h^*)\left(\left(\frac{\partial\vartheta_x^*}{\partial x^*}\right)^2 + \left(\frac{\partial\vartheta_y^*}{\partial y^*}\right)^2\right) - \left(\nu_1^*(h^*) + \nu_2^*(h^*)\right)\rho^*(h^*)\{\vartheta_x^*, \vartheta_y^*\}\right)$$

where $\qquad F^* = F/\rho_c\vartheta_c^2.$ (23)

Thus, the thermodynamic approach allows us the present of the Navier-Stokes equations for incompressible fluid in a compact form:

$$\frac{\partial \vec{\vartheta}^*}{\partial t} = -\frac{1}{\rho^*(h)}\left(\frac{\partial \dot{\vec{F}}^*(h)}{\partial \vartheta^*}\right), \quad \text{or} \quad \frac{\partial \vec{\vartheta}^*}{\partial t} = \frac{1}{\rho^*(h)}\left(\frac{\partial \dot{\vec{S}}^*(h)}{\partial \vartheta^*}\right).$$

**6. The tornado internal energy.** The law of free energy conservation (1) in dimensionless form is

$$\frac{dF^*}{dt} = -\left(\sigma^{e*} + \sigma^{i*}\right), \quad \text{or} \quad \frac{dF^*}{dt} = -\frac{dS^*}{dt}.$$

For a layer thickness *dh* and equal time intervals *dt*, we obtain an expression relating the reduced dimensionless free energy $dF^*$ with the reduced dimensionless entropy $dS^*$: $dF^* = -dS^*$. With the temperature changing $T_h$ relationship between the contribution of the internal energy *U* and entropy *S* changes according to the thermodynamic expression $F = U - T_h S$. Making the variables change $F^* = F/\rho_c \vartheta_c^2, S^* = ST_h/\rho_c \vartheta_c^2, U^* = U/\rho_c \vartheta_c^2$, we come to the dimensionless thermodynamic equation $F^* = U^* - S^*$. For a layer thickness *dh* last expression can be rewritten $dF^* = dU^* - dS^*$. Comparing this expression with $dF^* = -dS^*$, we conclude that $dU^* = 0$. Thus, for a certain layer *dh* internal energy is constant (time independent), since the model does not take into account phase transformations.

### 4. Formulation of the boundary problem

Equation (16) with initial and boundary conditions represent the boundary problem for a horizontal layer of height *dh*. The problem is to find solutions of the equation with the definition of the velocity components projections $\vartheta_x^*(x^*, y^*, t), \vartheta_y^*(x^*, y^*, t)$ onto the *xy* plane and to identify all functions that depend on them: $\vartheta^*$ – speed module; $p^*$ – pressure; $|\vec{\nabla}p^*|$ – the module of pressure gradient; $\sigma^{i*}$ – the entropy production; $\sigma^{e*}$ – reversible flows of entropy; $\dot{F}^*$ – free energy change rate; $\dot{S}^*$ – the total entropy change rate.

**Parameters of the problem.** The problem is multiparametric. Were introduced following parameters of the problem: $\ell_0$ – the spatial scale (*m*) (the tornado size – size of large vortex at the surface (diameter)); $p_c, \rho_c$ – pressure and air density at sea level ($Pa/m^2$, $kg/m^3$), $t_0$ – the time scale (*s*); $\vartheta_c$ – velocity scale, defined by lateral wind speed at ground level ($[\vartheta_c] = m/s$); $T_0 = T_h$ – an average layer homogeneous temperature at the altitude *h* (*K*); $h_0$- maximum height of the tornado arm (*m*); $2r_0^k$ – diameter of the core (arm) at the altitude *h* (*m*); *q* – the quantity characterizes strength of movement sources ($[q] = 1/s$), the initial choice of which was made to obtain spiral waves solutions; $\hat{\alpha}$ – tensor characterizing the sinks ($[\alpha_i] = s/m^2$), which was chosen in accordance with the value of the parameter *q*, $\mu$ – the molar mass of air (*kg/mol*), $\hat{\eta}$ – the tensor describing the viscosity (*Pa·s*), $\tau$ – stress relaxation time (*s*).

The following reduced equation parameters are specified. According to the model kinematic viscosity considered as altitude independent: $v^*_1 = v^*_2 = v^*(h)$ (see Table 2 in Appendix 2). Sources and sinks function parameters were chosen constant: $q^* = 1$, $\alpha^*_1 = 10$, $\alpha^*_2 = 1$, $c_2 = 0.1$, the relative magnitude of the density $\rho^*$ and pressure $p^*$ depending on the altitude *h* for the analyzed layer shown in Table 4 (Appendix 2).

To avoid possible discrepancies chosen design scheme for the time step in the range of $2.5*10^{-7} \cdot t_0$ to $5*10^{-7} \cdot t_0$ the size of the area was defined as 500x500 calculating points with *x* and *y* steps:

$\Delta_x = \Delta_y = 0.002$, i.e. $x^* \in [0,1]$, $y^* \in [0,1]$. The transformation to real space coordinates carries out by formula: $y = \ell_0 y^s / 500$, $x = \ell_0 x^s / 500$, where $x^s, y^s$ – the coordinates of the spatial grid.

**1. Initial conditions.** For the beginning of rising air spinning provision a side wind with speed increases with altitude is needed. Let this velocity at a distance $\ell_0/2$ from the center of the tornado equal $\vartheta_c$ on the surface, then the variable in the Kuramoto-Tsuzuki equation (16) is $\Phi = \vartheta_x^* + i\vartheta_y^*$, where $\vartheta_x^* = \vartheta_x/\vartheta_c$, $\vartheta_y^* = \vartheta_y/\vartheta_c$ – dimensionless projections of the velocities to the corresponding axis. Therefore, the initial conditions for the function $\Phi(x^*,y^*,t^*)$ were specified as follows:

$$\Phi(x^*,y^*,0) = \Phi_0 \exp\left(i \cdot \left(R_0 \sqrt{(x^* - x_c^*)^2 + (y^* - y_c^*)^2} + m\varphi - \frac{\pi}{2}\right)\right), \text{ where } \varphi = arctg\left(\frac{y^*}{x^*}\right);$$

here $\Phi_0$ – velocity amplitude, $R_0$ – a wave vector module along the radius vector, $m$ – the topological charge, $x^* = x/\ell_0$, $y^* = y/\ell_0$, $x_c^* = x_c/\ell_0$, $y_c^* = y_c/\ell_0$. Initially for numerical calculations as the initial conditions (initial perturbation) at all altitudes the left or right twist of the spiral wave were set with the topological charge equal to $m = -1$ (two-arm wave) around the area center with coordinates $x_c^* = 0.5$, $y_c^* = 0.5$. This means that the tornado center is in the middle of the computational area. Other parameters are initially chosen to be iquall to: $\Phi_0 = 10$, $R_0 = 60$. The value of $m$ we specify, as +1, and −1, which corresponds to the right and left tornado twist, respectively. Question about tornado twisting associated with the winds dynamics at the time of its formation. In areas where the wind rose is constant, we should expect the appearance of only one twist. However, where the wind rose is changing, twist can change too. The necessity to consider a few turns of the spiral wave for tornado core formation visual simulation are explain high value of the coefficient $R_0$. The velocity amplitudes $\Phi_0$ were chosen from the condition (22) which necessary for the emergence and development of self-organization in such structure. In general, the solution procedure contained a definition of extreme motion characteristics dependence on the values of $\Phi_0, R_0$ and $m$.

**2. Boundary conditions** were set as follows:

$$\Phi(0,y^*,t^*) = 0 \qquad \Phi(1,y^*,t^*) = 0 \qquad \dot\Phi(0,y^*,t^*) = 0 \qquad \dot\Phi(1,y^*,t^*) = 0$$
$$\Phi(x^*,0,t^*) = 0 \,, \qquad \Phi(x^*,1,t^*) = 0 \,, \qquad \dot\Phi(x^*,0,t^*) = 0 \,, \qquad \dot\Phi(x^*,1,t^*) = 0 \,.$$

As follows from the above, two-arm spiral waves with the left twist ($m = -1$) are among the self-similar solutions of equation (16). But because of the numerical calculation time length to simulate the tornado development it is better to choose the initial conditions close to the asymptotically stable, so that less time waiting for the numerical solution asymptotic "exit" to the nearest self-similar, that was done. It is assumed that if the regime is asymptotically stable, then it is stationary.

Equation (16) was solved by the method of alternating directions numerically, which consists of dividing the time step $\Delta_t$ into the number of independent spatial variables – in this case into two. At each fractional time step one of the spatial differential operators is explicitly approximated, while the other – implicitly. At the next fractional time step sequence of approximation is changed.

## 5. Calculation results. Layerwise spatial and force characteristics of the pressure in the tornado

This section presents the results obtained by numerical methods for the model constructed in sections 1-3 and of the boundary problem, described in Section 4. For the calculation points grid $(x^s, y^s)$, where $x^s \in [0,500]$ and $y^s \in [0,500]$ – coordinates of the chosen spatial grid, the components of velocity $\vartheta_x^*(x^s, y^s), \vartheta_y^*(x^s, y^s)$ in a flat layer whip-like tornado was identified. The transformation to real spatial coordinates is carried out here and later by the formulas: $y = \ell_0 y^S / 500$, $x = \ell_0 x^S / 500$. By found components f the velocity vector vector-field of airflow velocity in a tornado are built and velocity module of the airflow $\vartheta^*$, the maximum $\vartheta_{max}^*$ and minimum $\vartheta_{min}^*$ values were found.

As a consequence of problems with the sharpening spatial characteristics of the core formation, the diameter of the self-organization area $d_0$, which also corresponds to a zone of reduced pressure, the width of the pressure equalization ring to atmospheric pressure at the edge of the basin, and pressure characteristics, including pressure gradients were identified, which makes such an approach different from others.

Two cases for finding the pressure gradient and pressure in the tornado flat layer was considered: non-potential (Part 1) and the potential flow in terms of physically understandable Bernoulli equation (part 2). For these cases some of the spatial tornado characteristics evaluated.

The last part presents a comparative analysis of the results obtained by two methods, which provides non-triviality of the chosen approach.

### 5.1. The case of non-potential flows

Consider the more general case of having the function of movement sources and sinks in the tornado, for which in the equation for nonpotential flow (16) constant $q$ and determinant $\hat{\alpha}$ different from zero. In this case to determine the pressure gradient in the tornado we need to use equation (11), taking into account the existence of the function of sources and sinks.

### 5.1.1. Magnitude of velocity for non-potential flow. The occurrence of localized structures. The core of the tornado

On Fig. 2 on the basis of numerical solutions for non-potential flow equation (16) with nonzero values of the function of sources and sinks the dimensionless velocity module is represented as a function of the coordinates $x_s, y_s$ – the dimensionless coordinates of the grid in the central tornado part. Module of the velocity $\vartheta^* = \sqrt{\vartheta_x^{*2} + \vartheta_y^{*2}}$ depends on initial conditions and varies of values from 0 to 5.7 ($\Phi_0 = 10, R_0 = 60, m = -1$), having at first glance, the strange form. In the diagonal section two distinct maximum of velocity can be seen. The highest speed attained for values of the dimensionless radius $r_0^* \equiv r_0 / \ell_0 \approx 0.048$. In the layer the horizontal velocity components was calculated, which are then used to calculation other characteristics.

Sharpening regime (blow-up regime) in a nonlinear medium is generated by intrinsic nonlinearity of the medium. For this environment in a nonlinear way the tornado sources of momentum (speed) must work, what can be seen from equations (15). In this case the competition between increment process (source) and dissemination (the viscous properties) of the pulse gives to a new medium characteristics appearance: a linear dimension on which these processes "balance" each other [5]. Momentum source in the medium leads to a sharpening regime and the development of this mode leads to the appearance of localized complex structures. The source in this medium is a complex self-organized system of vortices prior to the tornado formation.

Fig. 2 shows the estimation formulas for the spatial characteristics of the core relative to the tornado characteristic size $\ell_0$, about the thermodynamic determining method of which will be discussed later. Description of characteristics is given in Table 1.

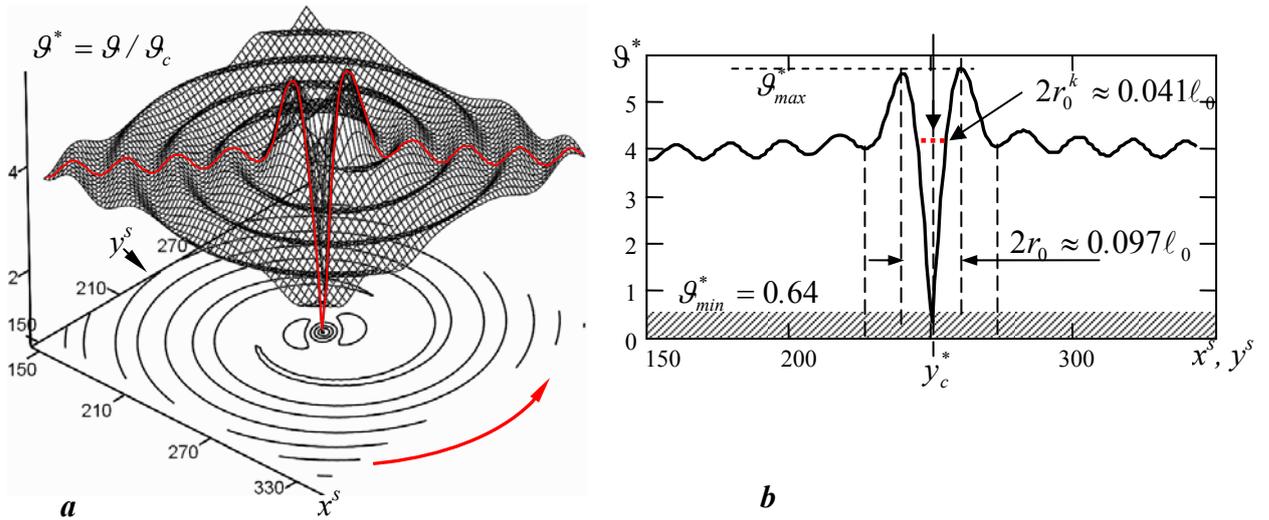

Fig. 2. Magnitude of horizontal velocity $\vartheta^*$ for nonpotential flow in the presence of sources and sinks for a flat layer in the steady tornado state near the ground $h \approx 0$ m and the center of the region (top view) (*a*) and its diagonal cross section (*b*). The size of the given area is 200x200 point calculated by the total number of points calculated of 500x500. Red arrow pointed the direction of twist.

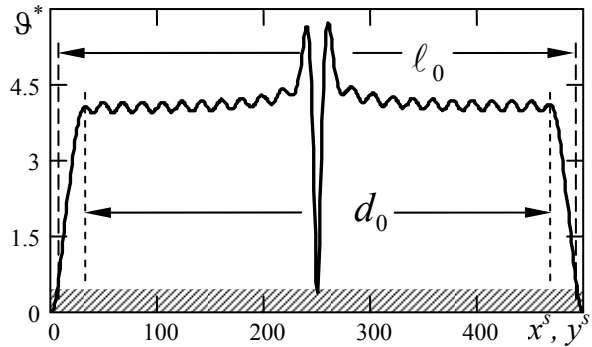

Fig. 3. The diagonal section of the horizontal velocity module $\vartheta^*$ for non-potential flow in the presence of sources and sinks in the whole flat layer tornado region (500*500 points) in the steady state near the ground for $h \approx 0$ m at time $t=2*10^{-4}t_0$. Spatial step $\Delta_x=\Delta_y=0.002$. $d_0$ – low-pressure zone diameter (self-organization area), $\ell_0$ – the diameter of the tornado basin.

Determination $\ell_0$ was carried out for the number of points calculated, which varies from 500 to 600. Boundary and initial conditions at the same time remained constant. The average value $\ell_0$ was calculated as $\ell_0 = 500 \pm 7$ points. It should be noted that the area size with the number of calculation points increasing grows, starting at 530. The maximum and minimum speed values are not changed, indicating the weak influence of boundaries on the structure occurred (the core) stability in the tornado center.

In the numerical calculations the unusual behavior of the velocity near the center of the field (Fig. 2.3) documented. We can assume that the occurrence of such stationary core in a tornado plane layer – a localized area in the figure – is the result of sharpening in the effort of equation (16) solution to the self-similar solution of the spiral wave. Real tornado in this case goes to the stationary state and the core is formed into a flat layer, whose size can be numerically defined.

Table 1

**Tornado spatial characteristics for nonpotential flow, found by velocity magnitude in the presence of sources and sinks,** $q^*=1$, $\alpha^*_1=10$, $\alpha^*_2=1$, $\Phi_0=10, R_0=60, m=\pm1$.

| Basin diameter, m | $\ell_0$ | 500 | 1000 | 1500 | 2000 |
|---|---|---|---|---|---|
| Basin square, $km^2$ | $\pi\ell_0^2/4$ | 0.196 | 0.785 | 1.767 | 3.14 |
| Inner core diameter, m | $2r_0^k \cong 0.022\ell_0$ | 11 | 22 | 33 | 44 |
| Outer core diameter, m | $2r_0 \cong 0.04\ell_0$ | 20 | 40 | 60 | 80 |
| Perimeter of circle, m | $\pi\ell_0$ | 1570 | 3140 | 4710 | 6280 |
| Low-pressure and self-organization zone diameter, m | $d_0 \cong 0.878\ell_0$ | 439 | 968 | 1317 | 1756 |
| Width of the pressure equalization ring to atmospheric pressure $p_c$ | $\Delta_c = \frac{(\ell_0 - d_0)}{2}$<br>$\Delta_c \cong 0.06\ell_0$ | 30 | 60 | 90 | 120 |

In the core center ($x^s=y^s=250$) there is no horizontal movement of air. Velocity magnitude increases with distance from the center and approaches to the core radius. Relative maximum velocity of rotation $\vartheta^*_{max}$ is observed along the core radius around which two mesovortices defining this core twist (Fig. 2b). Figure below (Fig. 2a) shows a velocity projection (top view).

The upper boundary of the shaded region in Fig. 2 corresponds to a lower condition of self-organization in the presence of sources and sinks function for the chosen parameters $q^*=1$, $\alpha^*_1=10$, $\alpha^*_2=1$ ($\Phi_0=10$ $R_0=60$, $m=-1$). Thus, for non-potential flows in these parameters self-organization (22) holds in the velocity range from $\vartheta^*_{min}=0.64$ to $\vartheta^*_{max} \approx 5.7$.

We call the spatial region bounded by the outer $2r_0^*$ and inner $2r_0^{k*}$ diameter, which has the highest values of the horizontal rate, wall thickness of the *tornado core*. It is assumed that the visible diameter of a tornado is $2r_0^{k*}$, if different objects, trees, sand, etc. don't fall into tornado. Otherwise, the visible diameter is $2r_0^*$. For a closed streamline these areas with peaks of speed rotate along the spiral wave and create conditions for the existence of area with increased velocity. As you know, the definition of air velocity in the funnel is still a serious problem, including due consideration of only the idealized case of potential flows. Figure 2 shows an unwinding spiral wave for nonpotential flow in the region outside the core, and we can determine the relative velocity module as a function of relative spatial coordinates $x^s, y^s$. Near the land surface its diameter is only about 100 *m* [17].

**Influence of the amplitude $\Phi_0$ and $R_0$ to the solution of (16).** To determine the correct choice of initial conditions research of relationship between $\vartheta^*_{max}$ and the average velocity $\bar{\vartheta}^*$ over the layer of the amplitude $\Phi_0$ was carried out at a fixed value of the modulus of the wave vector along the radius vector $R_0$ and *m* (Fig. 4a, b).

With velocity amplitude $\Phi_0$ increasing, at fixed values $R_0=60$ and $m=-1$, maximum $\vartheta^*_{max}$ and average speed $\bar{\vartheta}^*$ on the field increased by a cubic polynomial. For certain values $\Phi_0$ the magnitude $\vartheta^*_{max}$ and $\bar{\vartheta}^*$ according to this approximation (Fig. 4a, b – solid line) are zero. The data were approximated by the formulas $\vartheta^*_{max} = 0.0141 \cdot (\Phi_0)^3 - 0.3583(\Phi_0)^2 + 3.2661\Phi_0 - 5.3019$, $\bar{\vartheta}^* = 0.0067 \cdot (\Phi_0)^3 - 0.1918(\Phi_0)^2 + 1.9415\Phi_0 - 2.8464$ with the reliability of approximation $M^2=1$. The range of speed, for which the conditions of self-organization (22) ($\vartheta^* \leq 0.64$) not hold and the limiting minimum value $\Phi_0^{min}$ marked which below self-organization does not arise, shaded.

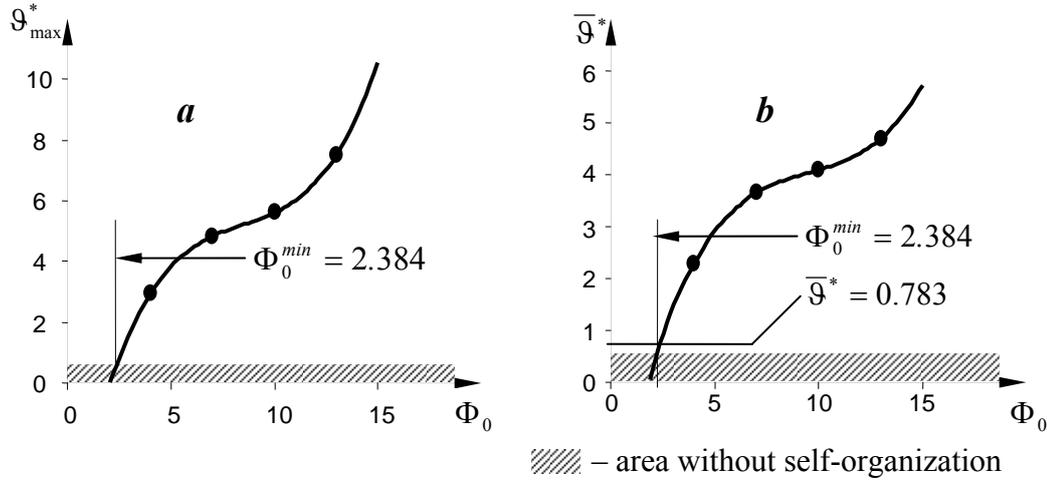

Fig. 4. The dependence of the maximum horizontal tornado speed $\vartheta_{max}^* = \vartheta_{max}/\vartheta_c$ in a layer at a height $h = 0$ m from the velocity amplitude $\Phi_0$ in the initial conditions (*a*) and average speed $\overline{\vartheta}^* = \overline{\vartheta}/\vartheta_c$ for non-potential flow around the basin of tornado form velocity amplitude $\Phi_0$ (*b*) for fixed values of $R_0 = 60$ and $m = -1$ in case of non-potential flow.

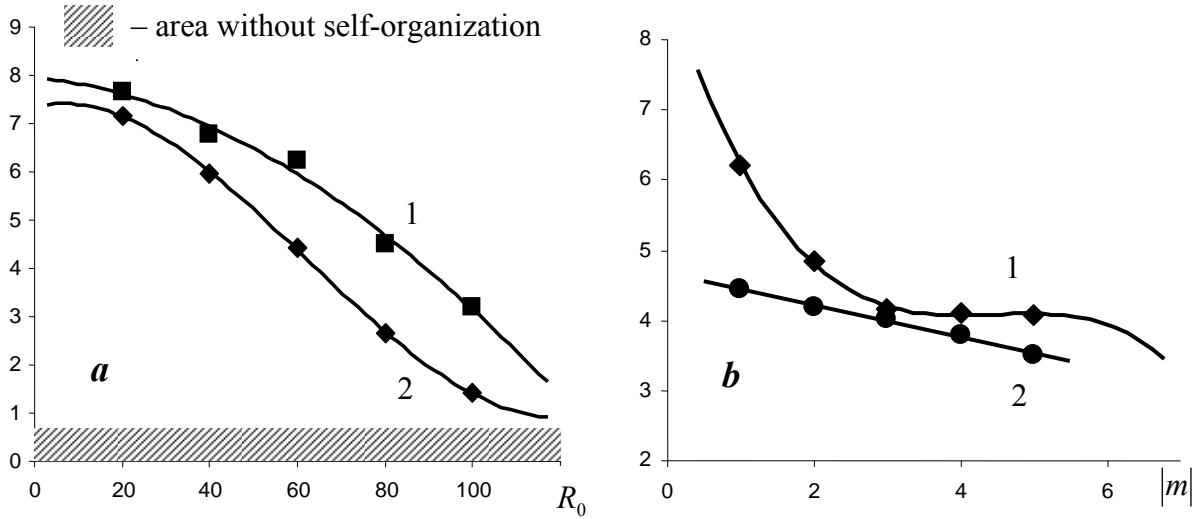

Fig. 5. The dependence to non-potential flow of the maximum horizontal tornado speed in a layer at a height $h=0$ m $\vartheta_{max}^* = \vartheta_{max}/\vartheta_c$ (curve 1) and average speed for the entire tornado basin $\overline{\vartheta}^* = \overline{\vartheta}/\vartheta_c$ (curve 2) from the value $R_0$ of fixed values $\Phi_0 = 10$ and $m = -1$ (*a*) and modulus from the topological charge $|m|$ (*b*) in fixed value $\Phi_0 = 10$ and $R_0 = 60$.

Additionally different values of a radial part $R_0$ considered. The data were approximated by the following functions (solid lines): $\vartheta_{max}^* = 4\cdot 10^{-7}(R_0)^3 - 0.0004(R_0)^2 - 0.0085 R_0 + 7.9494$, $\overline{\vartheta}^* = 9\cdot 10^{-6}(R_0)^3 - 0.0017(R_0)^2 + 0.0198 R_0 + 7.3584$, ($M^2 = 1$) (Fig. 5*a*); $\vartheta_{max}^* = -0.055(|m|)^3 + 0.7352(|m|)^2 - 3.2408|m| + 8.791$, $\overline{\vartheta}^* = -0.2235|m| + 4.6569$ ($M^2 = 1$), (Fig. 5b). The range of speeds for which the self-organization conditions (22) ($\vartheta^* \leq 0.64$) not hold is shaded.

According to Fig. 5a, for fixed values of $\Phi_0 = 10$ and $m=-1$, with increasing $R_0$ values of $\vartheta^*_{max}$ and $\bar{\vartheta}^*$ decreasing and may extend to values close to saturation at a certain $R_0$. Consequently, large amplitude of rotation speeds in a tornado funnel remain in a relatively small number of turns of the spiral wave. Fig. 5b shows that the speeds $\vartheta^*_{max}$ and $\bar{\vartheta}^*$ decrease with increasing modulus of the topological charge $|m|$. In the formation of additional arms in the tornado maximum and average speeds are reduced, and, according to the approximation, can take zero values.

The existence of extreme values of the speed with sharpening also leads to extreme values of the pressure gradient.

### 5.1.2. The pressure gradient in the layer of the tornado basin for non-potential flow

To determine the pressure gradient in the case of nonpotential flow in the presence of the function of sources and sinks is necessary to make the calculation according to expression (11). The numerical results for the calculation of the pressure gradient for that flow near the surface at height $h = 0$ m are shown in Fig. 6.

At the tornado core with the maximum speed there is extremely low pressure area, where the dust is drawn, forming a dark arm rising to the sky. At the same time in the center ($x^s= y^s=250$), as follows from the figure, it is no rotation, the pressure gradient is close to zero. The pressure is the lowest there due to centrifugal forces that tend to "push" the air closer to the periphery.

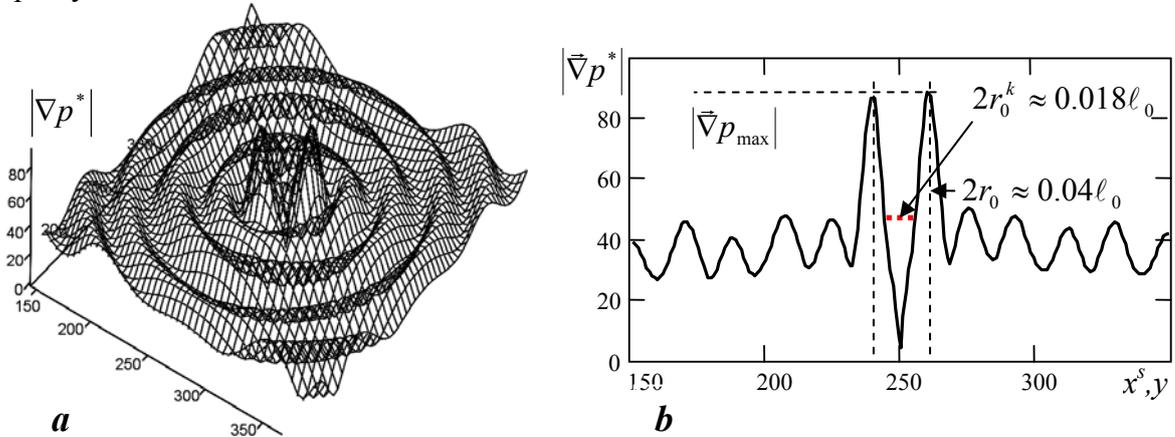

Fig. 6. Module of the pressure gradient $|\vec{\nabla}p^*|$ for non-potential flow in the presence of movement sources and sinks in the middle area of tornado (in a plane layer) near the ground $h \approx 0$ m near the center of the field (a) and its diagonal cross section (b). The size of the reduced area of 200x200 points calculated by the total number of 500x500 points calculated.

This causes the direction of the pressure gradient from the center of (increasing the pressure gradient up to large values ($|\vec{\nabla}p^*|>80$) during the transition from center to the boundary of the tornado core (Fig. 7a). Perhaps this is connected with the fact that around the pipe of the rising air in the center very strong wind blows, uproot the plants and destroying buildings. Module of the pressure gradient is maximum in the transition to the radius of the core $r < r_0^k$, in which there is a relative maximum level rotation speed $\vartheta^*_{max} \approx 5,7$. Around this core two mesovortices are twisted (b), which determine core dynamics. In the Table 2, in the basis on numerical experiments for the tornado of various power spatial characteristics of the core are defined, the diameter of the low pressure zone, as well as the width of the basin edge ring, at which the pressure becomes atmospheric. The definition of close observation at least one of characteristics can immediately appreciate the rest.

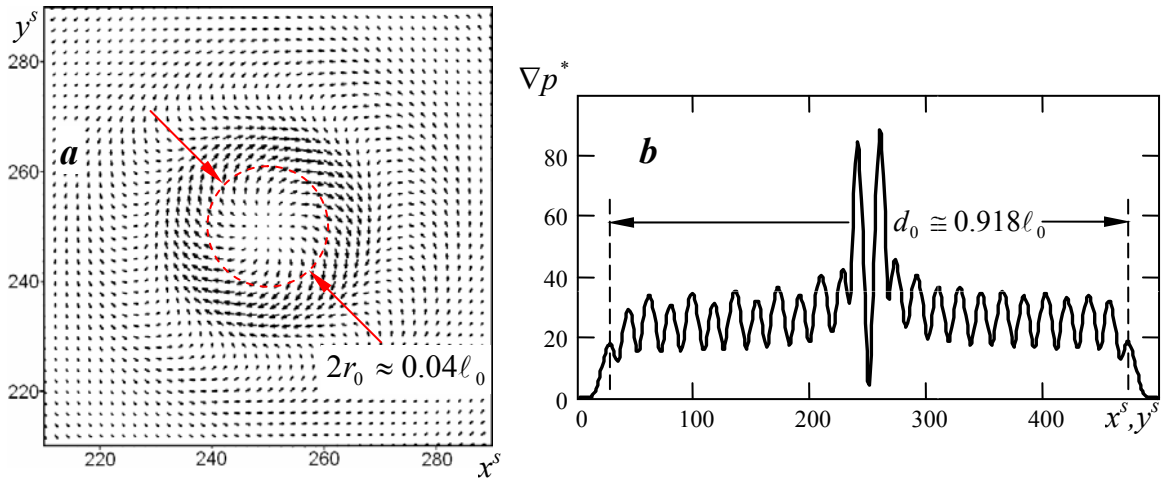

Fig. 7. The pressure gradient for nonpotential flow in larger central region near the tornado core (the calculation of expression (11)), presented as a vector field (*a*) showing the spin, and the diagonal section of the module of the pressure gradient $|\nabla p^*|$ across the basin of the tornado (b). Characteristic transverse size of the tornado – the diameter $\ell_0$ is calculated 500x500 points.

Table 2

**Tornado spatial characteristics for nonpotential flow, defined by the pressure gradient in to consider sources and sinks, $q^*=1$, $\alpha^*_1=10$, $\alpha^*_2=1$, $\Phi_0 = 10, R_0 = 60, m = \pm 1$.**

| Basin diameter, *m* | $\ell_0$ | 500 | 1000 | 1500 | 2000 |
|---|---|---|---|---|---|
| Basin square, $km^2$ | $\pi\ell_0^2/4$ | 0.196 | 0.785 | 1.767 | 3.14 |
| Inner core diameter, *m* | $2r_0^k \cong 0.018\ell_0$ | 9 | 18 | 27 | 36 |
| Outer core diameter, *m* | $2r_0 \cong 0.04\ell_0$ | 20 | 40 | 60 | 80 |
| Perimeter of circle, *m* | $\pi\ell_0$ | 1570 | 3140 | 4710 | 6280 |
| Low-pressure and self-organization zone diameter, *m* | $d_0 \cong 0.918\ell_0$ | 459 | 918 | 1377 | 1836 |
| Width of the pressure equalization ring to atmospheric pressure $p_c$ | $\Delta_c = \frac{(\ell_0 - d_0)}{2}$ $\Delta_c \cong 0.04\ell_0$ | 20 | 40 | 60 | 80 |

The pressure can be calculated for non-potential flows only through the integration of the equation (11). Therefore, if we use the potential approximation – the law of Bernoulli pressure (see next section), we can estimate not only the pressure gradients, but also the pressure itself.

### 5.2. Potential flows approximation.
### The pressure in the localized structures

If you exclude from consideration the function of sources and sinks the expression (12) for nonpotential flow reduces to potential flow case – Bernoulli's law (13), which can be used to estimate the pressure and pressure gradient in the tornado, which greatly simplifies the problem making it more physical. In this case, the velocity magnitude $\vartheta^*$ was found as before from the solution of the Kuramoto-Tsuzuki equation (16).

### 5.2.1. The pressure gradient in the basin of tornado in case of potential flow

Differentiating reduced pressure (13) to the coordinates $x^*$ and $y^*$ we obtain the following expression for calculating the pressure gradient of the potential flow

$$-\vec{\nabla} p^* = \gamma \vartheta^* \vec{\nabla} \vartheta^*, \qquad \gamma = \frac{\rho \vartheta_c^2}{p_c}. \tag{24}$$

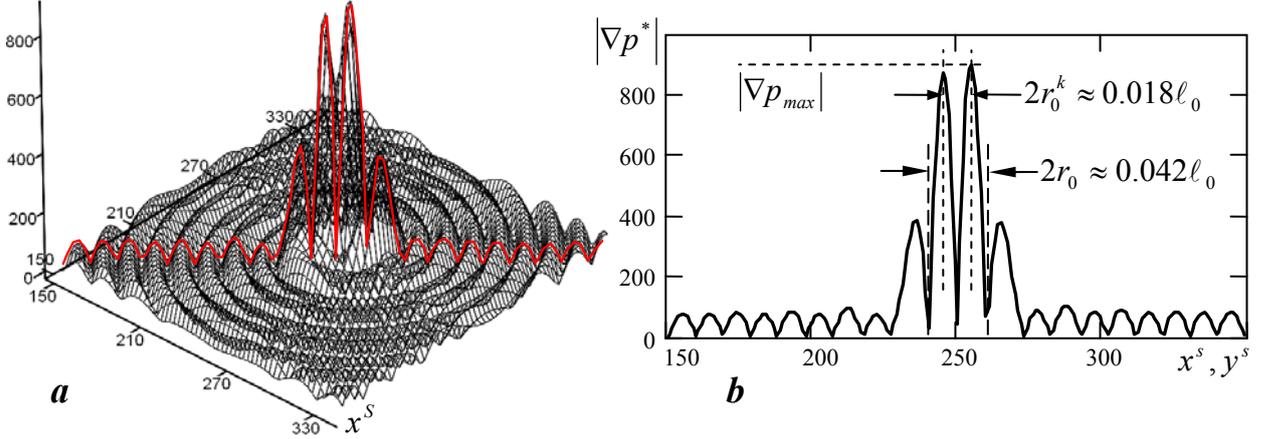

Fig. 8. Tornado middle area pressure gradient module $|\nabla p^*|$ in the stationary regime (in a plane layer) near the ground $h \approx 0$ m for the potential flow around the center region (a) and its diagonal cross section (b). The size of the reduced area of 200x200 points calculated at the full size 500x500.

Fig. 8 shows the modulus of the pressure gradient $|\nabla p^*|$ in the stationary regime in the middle of a tornado (in a plane layer) near the ground $h \approx 0$ m for the potential motion around the center region (a) and its diagonal cross section (b). The calculation was made according to formula (24).

In the center of the core ($x^s=y^s=250$) (Fig. 8,9) the pressure gradient is close to zero, as in the more general case of non-potential flows (Fig. 7), since the center is no horizontal movement of air (Fig. 9a and c). But in contrast to non-potential flows, the pressure gradient in the case of potential flow is strong bursts. Pressure gradient module is maximal for potential flow around a smaller radius, in which the relative maximum level speed of rotation $\vartheta^*_{max} \approx 5,7$ is observed along the radius of the core around which to spin two mesovortices.

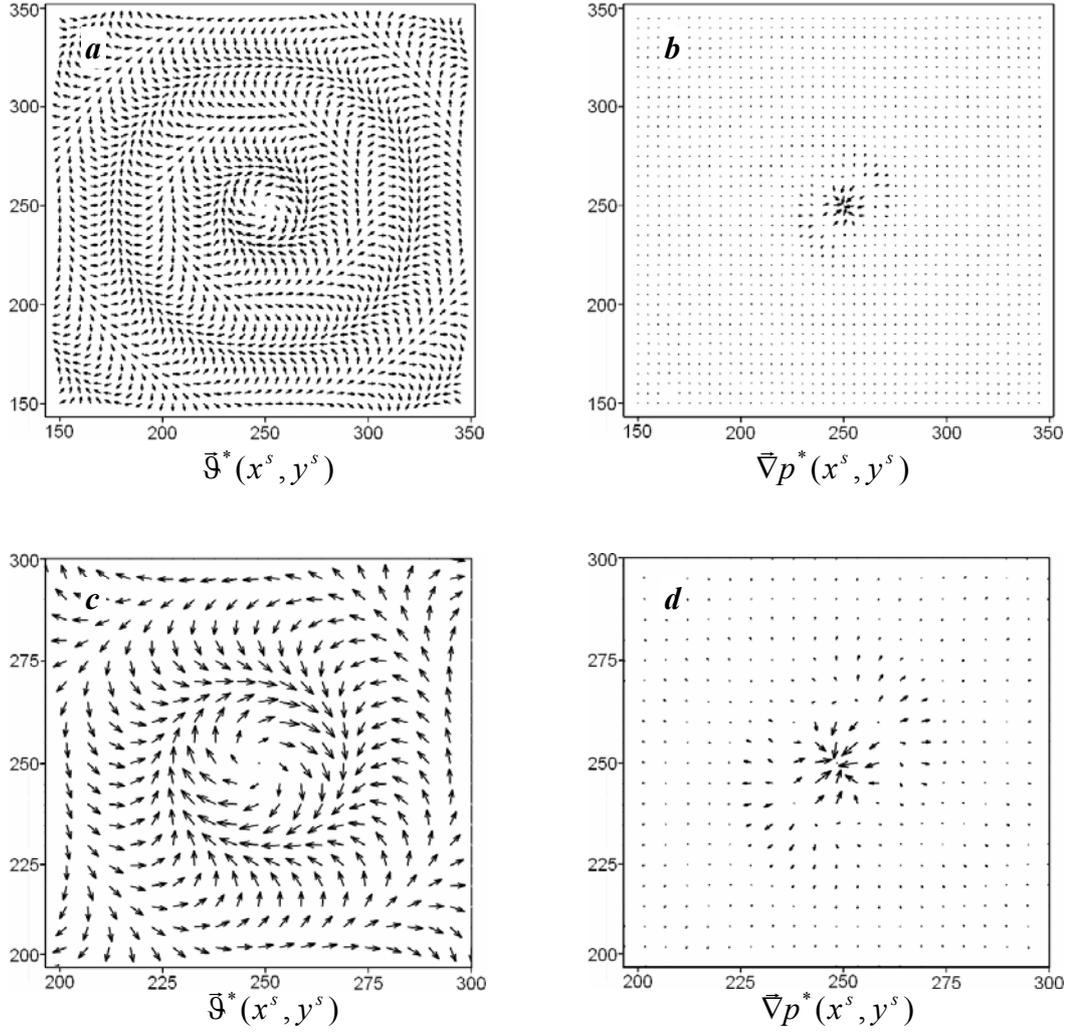

Fig. 9. Vector fields for the dimensionless modulus of the velocity $\vec{\vartheta}^*(x^*,y^*)$ (*a*) and reduced the pressure gradient $\vec{\nabla}p^*(x^*,y^*)$ (*b*) for potential flow (in a plane layer) near the ground $h \approx 0\ m$ in the central region of the tornado 200*200 points. Vector fields for the reduced speed module $\vec{\vartheta}^*(x^*,y^*)$ (*c*) and the reduced pressure gradient $\vec{\nabla}p^*(x^*,y^*)$ (*d*) increased by 2 times the central region of the tornado (100x100 points).

### 5.2.2. The pressure in the basin of tornado for potential flows

Transformation to the potential flow allows us to solve the inverse problem – to express the pressure as a function of speed. For non-potential flows, such a transformation is difficult. The pressure in the layer according to Bernoulli's law (13), assuming that the boundary pressure is equal to atmospheric pressure $p_c$, and boundary velocity of the flow is $\vartheta_c$ is to be determined by the following expression:

$$p^* = 1 + \frac{1}{2}\gamma(1 - \vartheta^{*2}),$$

The results are shown in Figures 10 and 11. Subject to $R_0$ pressures behave as follows (Fig. 10*b*): pressure $p^*_{max}$ does not depend on $R_0$ and is approximately equal to atmospheric pressure, $p^*_{min}$ and $\bar{p}^*$ increase and tend to the value of atmospheric pressure. Thus $p^*_{min}$ and $\bar{p}^*$ grow with the tornado number arms increasing.

The data were approximated by the following curves (solid lines) $\bar{p}^* = -5\cdot10^{-7}(R_0)^3 + 6\cdot10^{-5}(R_0)^2 + 0.0023R_0 + 0.6348$, $p^*_{min} = -2\cdot10^{-7}(R_0)^3 + 4\cdot10^{-5}(R_0)^2 + 0.0013R_0 + 0.6232$, $p^*_{max} = 1.006$, (Fig. 9a) $\bar{p}^* = 0.0102|m| + 0.8798$, $p^*_{max} = 1.006$, $p^*_{min} = 0.0044(|m|)^3 - 0.0551(|m|)^2 + 0.2265|m| + 0.6$ (Fig. 9b). Fig. 9b shows that the maximum pressure, as in the case shown in Fig. 5b, does not depend on the topological charge module and approximately equal to atmospheric pressure, although somewhat exaggerated. The reason for this excess is not entirely clear. What concerns $p^*_{min}$ and $\bar{p}^*$, then these values increase and tend to the value of atmospheric pressure. The case of leading center for m=0 was not considered due to a vortex-wave motion lack directed from the center (spin) in this case.

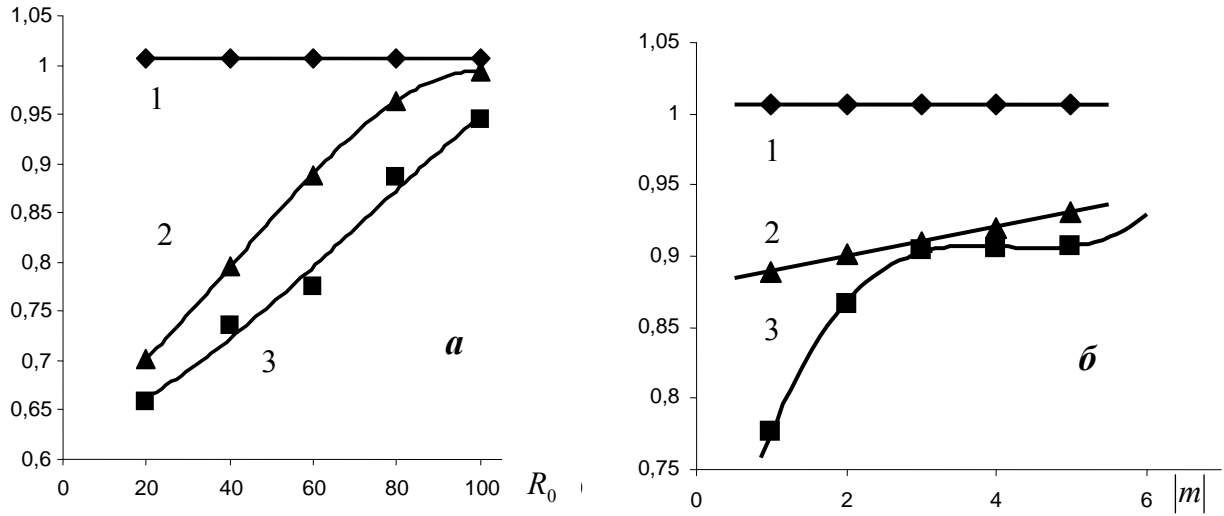

Fig. 10. Dependence of the maximum $p^*_{max}$ (curve 1), medium $\bar{p}^*$ (curve 2) and minimum $p^*_{min}$ (curve 3) pressures for the potential flow on the value $R_0$ with fixed values $\Phi_0 = 10$ and $m = -1$ (*a*) and on the magnitude $|m|$ with a fixed value of $\Phi_0 = 10$ and $R_0 = 60$.

**Twist mechanism.** On the basis of this result it is possible to reconstruct the following spinning mechanism for nonpotential and potential flows. Near the vortex pressure is strongly reduced due to high speeds, this causes the air in a layer several meters in height near the ground rush to the bottom of the vortex.

Reaching the edge the air begins to climb up the spiral until the top of the tornado does not merge with the air flows. As a result, most strong winds zone take the form of a closed ring (Fig. 2*a,b*), which is reflected in the figure. In the center the circles are visible – the tornado core (Fig. 2*b*), this is the closed circles (the layer is thin enough), also has a formidable structure. The vertical gradient of wind speed causes the rotation of air around a horizontal axis, and the presence of vertical motion of air causes it to spiral movement, which is reflected in Fig. 9 for the velocities and pressure gradients. Funnel through the pressure creates air movement. Arrows indicate the direction of air flow in areas of low and very low air pressure.

According to modern concepts, the formation of a tornado occurs in two stages [15]. At first, the entire column of rising air begins to twist, its diameter is of about 10-20 *km*, called mesocyclone. Mesocyclone with reduced pressure on its axis is like a vacuum cleaner hose. In the second stage, for reasons which, as noted in the literature [15], not yet understood, in mesocyclone closer to its periphery, a region with a diameter of usually not more than one kilometer arises, which at an altitude of several kilometers rotation is amplified. Then the rapid rotation is passed down, the vortex tube is stretched almost to the earth, "hanging" in only a few hundred feet above it. The core of the tornado forms, which is characterized by a specified size and speeds. This core has a thin wall – just a few meters (see Table. 3).

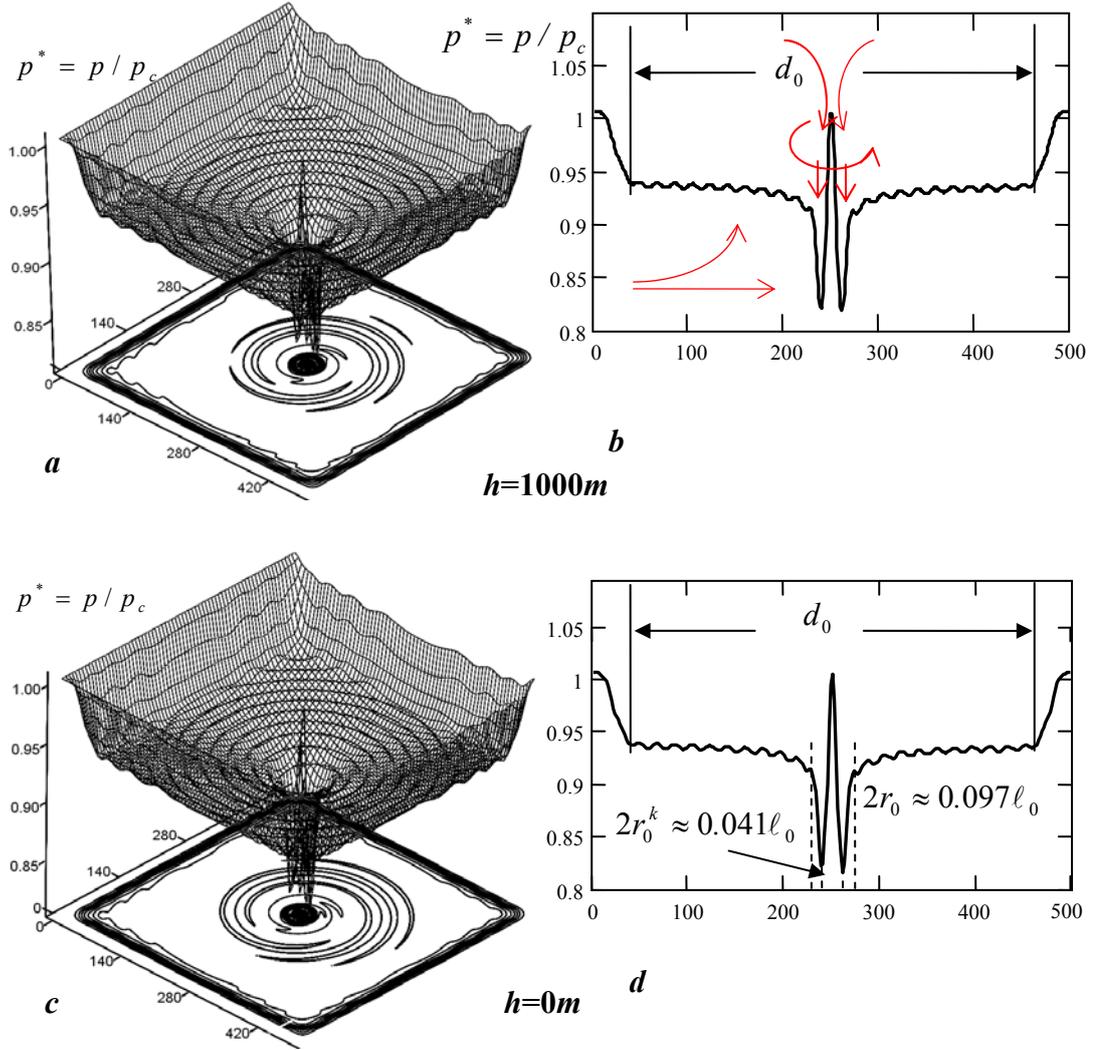

Fig. 11. The pressure in the local layers at the tornado total area for potential flow at altitude of $h = 1000$ and $0$ $m$ (*a*, *c*); bright tornado core revealed, its dimensions are given in Table 3 for the size $\ell_0$. The parameters $\vartheta_c = 22 m/s$, $\rho=1.087 kg/m^3$, $\nu=1.589*10^{-5} m^2/s$, $\rho^*=0.887$, $\nu_1^*=1.182$, $\nu_2^*=1.182$, $q^*=1$, $\alpha_1^*=10$, $\alpha_2^*=1$ at time $t=2*10^{-4} t_0$. The time step $\Delta_t=2.5*10^{-7}$; $\Delta_x=\Delta_y=0.002$. Area of 500 * 500 points calculated.

**Table 3**

**Spatial characteristics of whip-like tornado for potential flow, defined by the pressure at** $q^*=1$, $\alpha_1=10$, $\alpha_2=1$, $\Phi_0 = 10, R_0 = 60$, $m = \pm 1$

| | | | | | |
|---|---|---|---|---|---|
| Basin diameter, $m$ | $\ell_0$ | 500 | 1000 | 1500 | 2000 |
| Basin square, $km^2$ | $d_0 \cong 0.852\ell_0$ | 426 | 852 | 1278 | 1704 |
| Inner core diameter, $m$ | $\pi\ell_0^2/4$ | 0.2 | 0.785 | 1.767 | 3.14 |
| Outer core diameter, $m$ | $2r_0^k \cong 0.018\ell_0$ | 9 | 18 | 27 | 36 |
| Perimeter of circle, $m$ | $2r_0 \cong 0.042\ell_0$ | 21 | 42 | 63 | 84 |
| Low-pressure and self-organization zone diameter, $m$ | $\pi\ell_0$ | 1570 | 3140 | 4710 | 6280 |
| Width of the pressure equalization ring to atmospheric pressure $p_c$ | $\Delta_c = \dfrac{(\ell_0 - d_0)}{2}$ $\Delta_c \cong 0.07\ell_0$ | 37 | 74 | 111 | 148 |

On Fig. 11*b,d* tornado basin and its inner diameter $d_0$ is pointed, which are also determine the characteristic size $\ell_0$ and spatial characteristics of the core tornado. From Fig. 11 downward pressure phenomenon and the concept necessity for a tornado – the zone of lowered pressure – diameter $d_0$ corresponding to the self-organization area becomes understandable. Table 3 presents the cost formulas for the spatial characteristics of the tornado for potential flow, determined from numerical experiments. As expected, these results differ somewhat from the more accurate (see Table. 1,2), while yield values close to them.

Thereby it is proved that as a result of strong atmospheric pressure gradient in each layer occurring between the outer boundary of the tornado core and the surrounding area in the distance from the arm, the air abruptly changes direction and begins to gyrate up speed $\vartheta_z$ [15].

Core is a structure localized in the space that is stable in the period of the tornado existence. In the center of the tornado core pressure is equal to atmospheric pressure $p_c$ (Fig. 11). The lowest values are observed where the most maximum speed $\vartheta$. This is a plot of reduced pressure, and air rushes from the top down. At the tornado boundary, the pressure is also equal to atmospheric pressure. It is possible to reconstruct the mechanism connected with air suction, since the attractor basin at an altitude $h = 1000\ m$ is characterized by a lower pressure than the height $h = 0\ m$. The difference of pressures at two heights (see Fig. 12) is given by:

$$\Delta p(x^*, y^*) = p^*_{h=0}(x^*, y^*) \cdot p(0) - p^*_{h=1000}(x^*, y^*) \cdot p(1000) > 0,$$

where $p(h)$ – the dependence of the atmospheric pressure on the altitude $h$ (Table 2, Appendix 2). This causes the air movement upward instead of downward.

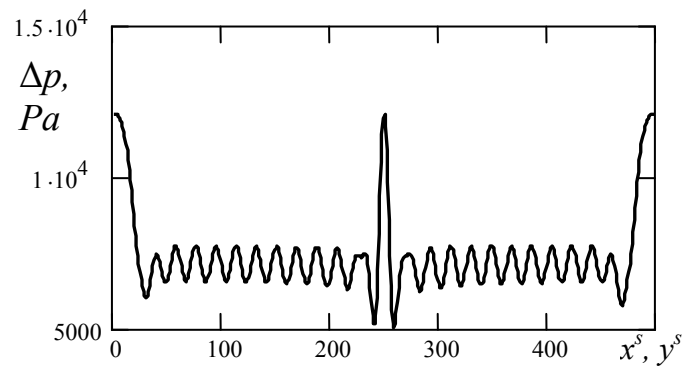

Fig. 12. The absolute pressure difference between two altitudes for the potential flow suggests that the pressure at the altitude $h = 1000\ m$ lower than at the height $h = 0\ m$. This causes the air movement upward instead of downward.

The air in the surface layer begins sucked into this mesocyclone. Form the top it goes to the core, where the zone of reduced pressure, from bottom and sides, which causes spin. Moreover, its horizontal velocity by a rotating column of up to 100-120 *km/h*.

The wavy surface of the pressure in the figures 11,12 related to the tornado spinning and caused by variable speed module in spirals. funnel – expanded core pressure for each layer – clearly drawn.

### 5.3. Comparison of tornado characteristics for nonpotential and potential flow

Table 4 compares the characteristics of a tornado, for nonpotential and potential flows. Estimate formulas for the spatial characteristics of tornado in the non-potential and potential flow is presented, determined from numerical experiments.

**Table 4**
**Determination of the whip-like tornado spatial characteristics
with $\Phi_0 = 10$, $R_0 = 60$ $\ell_0 = 500 м$, $|m| = 1$, a comparative table.**

| Characteristic | Flow type | Nonpotential flow | | | | Potential flow | |
|---|---|---|---|---|---|---|---|
| | | From velocities | | From pressure gradients | | From pressure | |
| Basin diameter, m | $\ell_0$ | 500 | | | | | |
| Low-pressure and self-organization zone diameter, m | $d_0$ | $d_0 \cong 0.88\ell_0$ | 439 | $d_0 \cong 0.92\ell_0$ | 459 | $d_0 \cong 0.85\ell_0$ | 426 |
| Inner core diameter, m | | $2r_0^k \cong 0.02\ell_0$ | 11 | $2r_0^k \cong 0.02\ell_0$ | 9 | $2r_0^k \cong 0.02\ell_0$ | 9 |
| Outer core diameter, m | | $2r_0 \cong 0.04\ell_0$ | 20 | $2r_0 \cong 0.04\ell_0$ | 20 | $2r_0 \cong 0.04\ell_0$ | 21 |
| Perimeter of circle, m | $\pi\ell_0$ | 1570 | | | | | |
| Width of the pressure equalization ring to atmospheric pressure $p_c$ | $\Delta_c = \frac{(\ell_0 - d_0)}{2}$ | $\Delta_c \cong 0.06\ell_0$ | 30 | $\Delta_c \cong 0.04\ell_0$ | 20 | $\Delta_c \cong 0.07\ell_0$ | 37 |

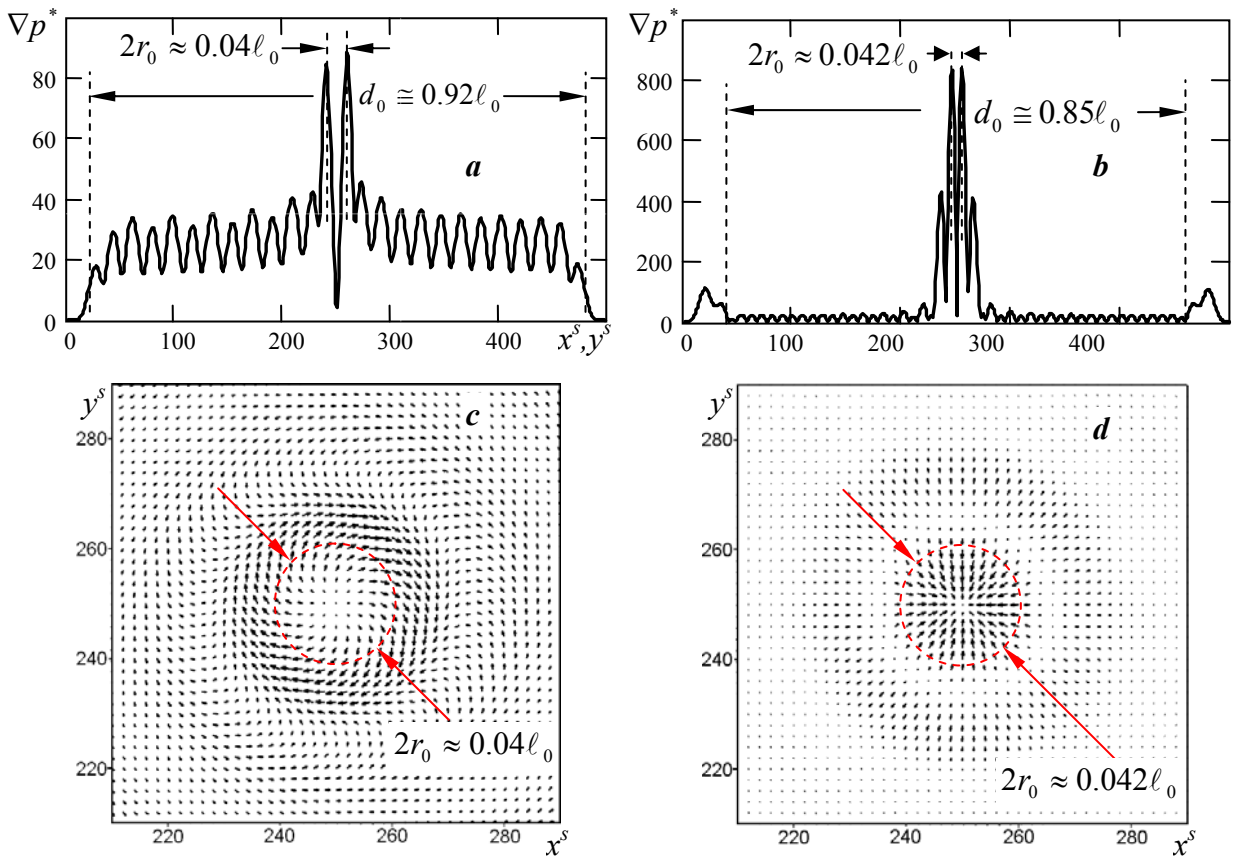

Fig. 13. The diagonal section of the pressure gradient module $|\nabla p^*|$ in steady state throughout the tornado basin (in a plane layer) at a height $h = 1000\ m$ and vector fields for the dimentionless modulus of the pressure gradient $\vec{\nabla} p^*(x^*, y^*)$ in the central region of the tornado (100x100 points) for nonpotential flow (*a* and *c*, respectively) and for the potential flow (*b* and *d*, respectively).

At Fig. 13 for comparison the diagonal sections of the pressure gradients modules $\left|\nabla p^*\right|$ in the steady state in the whole tornado basin (in a plane layer) at a height $h = 1000\ m$ and vector fields for the dimensionless modulus of the pressure gradient $\vec{\nabla} p^*(x^*, y^*)$ in the central region of the tornado (100x100 points) for nonpotential flow (*a* and *b*, respectively) and for potential flow (*b* and *d*, respectively) are presented. Note the different scale for the pressure gradient.

From the figures it follows that the calculation in the model of potential flow in the tornado core gives much higher values than the model for nonpotential flow including sources and sinks. This is explained by the fact that there are sinks in nonpotential flow model, while in the Bernoulli model they don't.

<h3 style="text-align:center">6. Calculation results. Thermodynamic<br>tornado characteristics</h3>

Thermodynamic approach, being an independent part of the article, make it possible to get a thermodynamic description of the tornadonon-linear physical effects observed and to explain nonlinear self-organization mechanisms in the tornado, by finding thermodynamic characteristics local to each layer at certain height – entropy production, change rate of the entropy and free energy. The approach not only complements conclusions in the previous section, but also gives new results.

### 6.1. The function of external sources and sinks. Spiral wave

Function of external sources and sinks is represented at Fig. 14. The calculation was based on the formula (10). External actions – the entropy flux

$$\sigma^{e*} = \frac{d_e S^*}{dt} \leq 0$$

in the central region at an height $h = 1000\ m$ is represented.

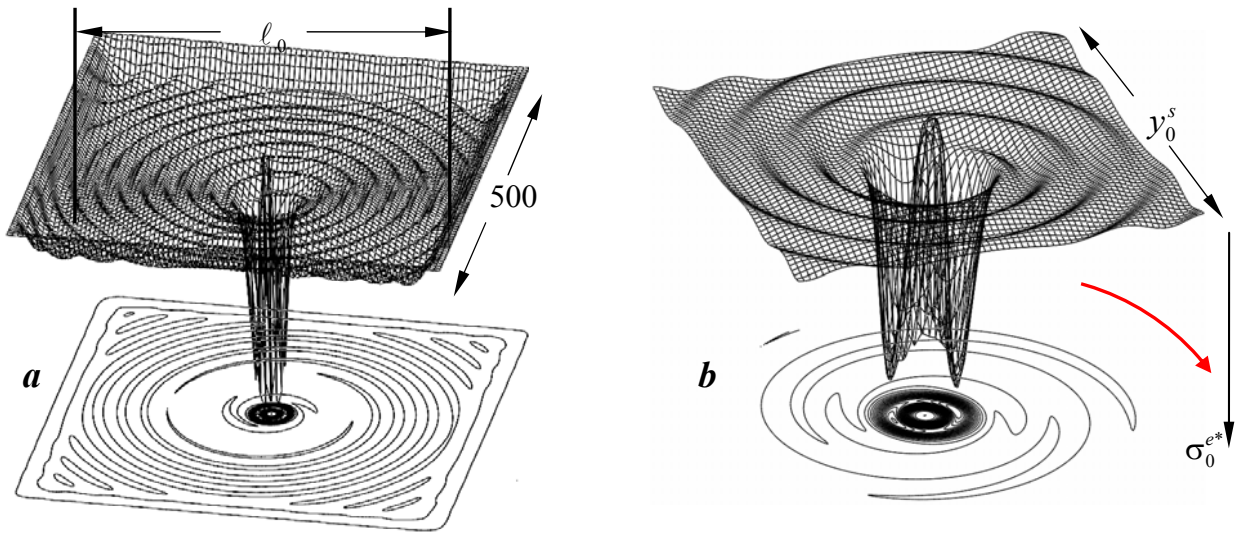

Fig. 14. External entropy flux and its projection bottom view for the plane layer in the tornado steady state (*a*), as the result of the sources and sinks function influence, this is function of spatial coordinates $x$, $y$ at the height $h = 1000\ m$; its central part scaling up (*b*). Parameters: $\rho = 1.087\ kg/m^3$, $\nu = 1.589*10^{-5}\ m^2/s$, $\rho^* = 0.887$, $\nu_1^* = 1.182$, $\nu_2^* = 1.182$, $q^* = 1$, $\alpha_1 = 10$, $\alpha_2 = 1$, at the time $\Delta_t = 2.5*10^{-7}$, $\Delta_x = \Delta_y = 0.002$, $t = 2*10^{-4} t_0$. $\sigma_0^{e*} = -2000$, $y_0^s = 200$.

The external entropy flux bottom view for a tornado plane layer (*a*), as the result of the sources and sinks effect, is represented. Immediately outside the core air rises with rapidly rotation, as a result a region of highly rarefied air is created. From Fig. 16 it follows that the spiral wave arises, which rotates counterclockwise when viewed from the top, and clockwise when viewed from below. This is clearly seen from the animation, which is constructed by numerical calculations. Changing the hyperbolic charge sign ($m = \pm 1$) can change the direction of rotation.

## 6.2. Entropy production.
## Creation of mesovortices as dissipative structures

Calculations by (17) showed that manyvortices structure is typical for tornado. This means that at the main tornado periphery air rotates around its axis ring-direction and simultaneously around the axis of the main tornado secondary vortices are formed.

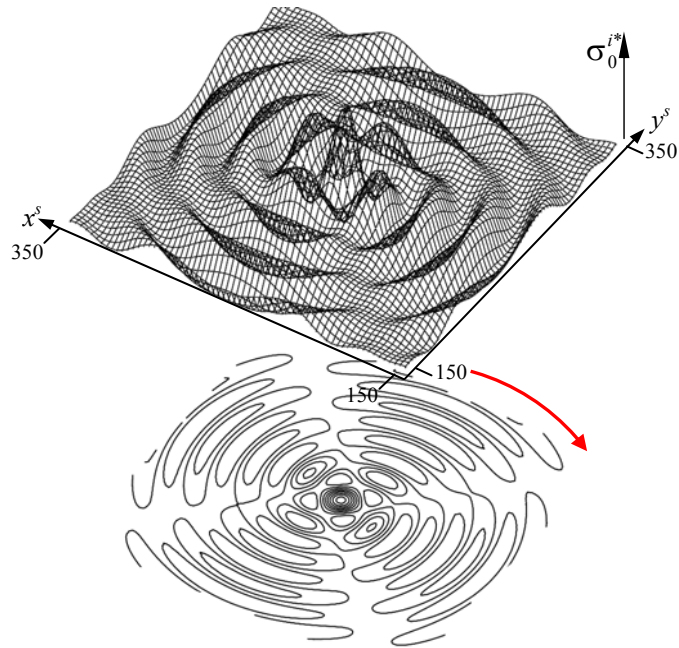

Fig. 15. Flat layer entropy production $\sigma^{i*}$ in tornado steady state at the altitude of 1,000 *m* as a function of spatial coordinates $x_s$, $y_s$. Mesovortices – ordered structures, as result of non-linear system response to external stimuli (external entropy flux) – is registered. Red arrow indicates the twist direction. Parameters: $\rho = 1.087\ kg/m^3$, $\nu = 1.589*10^{-5}\ m^2/s$, $\rho^* = 0.887$, $\nu_1^* = 1.182$, $\nu_2^* = 1.182$, $q^* = 1$, $\alpha_1 = 10$, $\alpha_2 = 1$, at the time $\Delta_t = 2.5*10^{-7}$, $\Delta_x = \Delta_y = 0.002$, $t = 2*10^{-4} t_0$. It is obtained that the entropy production in the core is maximal $\sigma_0^{i*} = 4 \cdot 10^4$.

The calculation was carried out under conditions (22). Entropy production

$$\sigma^{i*} = \frac{d_i S^*}{dt} \geq 0$$

indicates the emergence of dissipative structures as vortices (mesovortices) in the layer (Fig. 15), which coherently connected with each other, since the total entropy rate in the layer decreases. Their size is a wide range of certain values, and as noted in [12], can reach a few meters. However, large vortices comparable with the radius of the tornado $R_T = \ell_0 / 2$ can arise. Time dynamics of the vortices is probably responsible for the emerging multi-frequency rumble, extending from the arm in the space surrounding it. Namely this vortices cause strong winds and the most complicated cycloidal destruction traces. As a result, in some regions velocities added. For example, in the tornado path a house falls in that vortex just "explodes" from the powerful moments of the forces and pressures directed at the whole variety of possible directions.

Numerical simulation shows that if the conditions of self-organization (22) not hold, then the vortices disappear.

### 6.3. The total entropy change rate. The flow lines with an equal entropy change rate

The total entropy change rate was calculated by expression (21). Velocities interaction leads to an even smaller vortex structures formation than in the previous figure, which is reflected in the form of closed flow loops with an equal entropy change rate. The total entropy change decreases, because of the processes of self-organization.

From fig. 16 it follows that in each horizontal layer on a par with small mesovortices even smaller also ordered dissipative structures are fixed, that structures are non-linear system response to external stimuli. Their characteristics are as for the medium size vortices closed loops of the flow. In these structures the entropy decreases with time.

On the lower plane (Fig. 6) for a complete entropy change

$$\frac{dS^*}{dt} = \frac{d_e S^*}{dt} + \frac{d_i S^*}{dt}$$

the result of an interaction $\sigma^{e^*}$ and $\sigma^{i^*}$ is reflected, which confirms the emergence and development of "propeller" type dissipative structures, which also rotates along chosen initial conditions twist ($m = \pm 1$). As in the previous figure this is shows as formed by a given symmetry closed loops with a constant energy change rate. It is not excluded that the smallest vortices contain even smaller vortices, if we fixed other initial conditions and parameters of equation (16) – this is to talk about the fractality of that vortex formation.

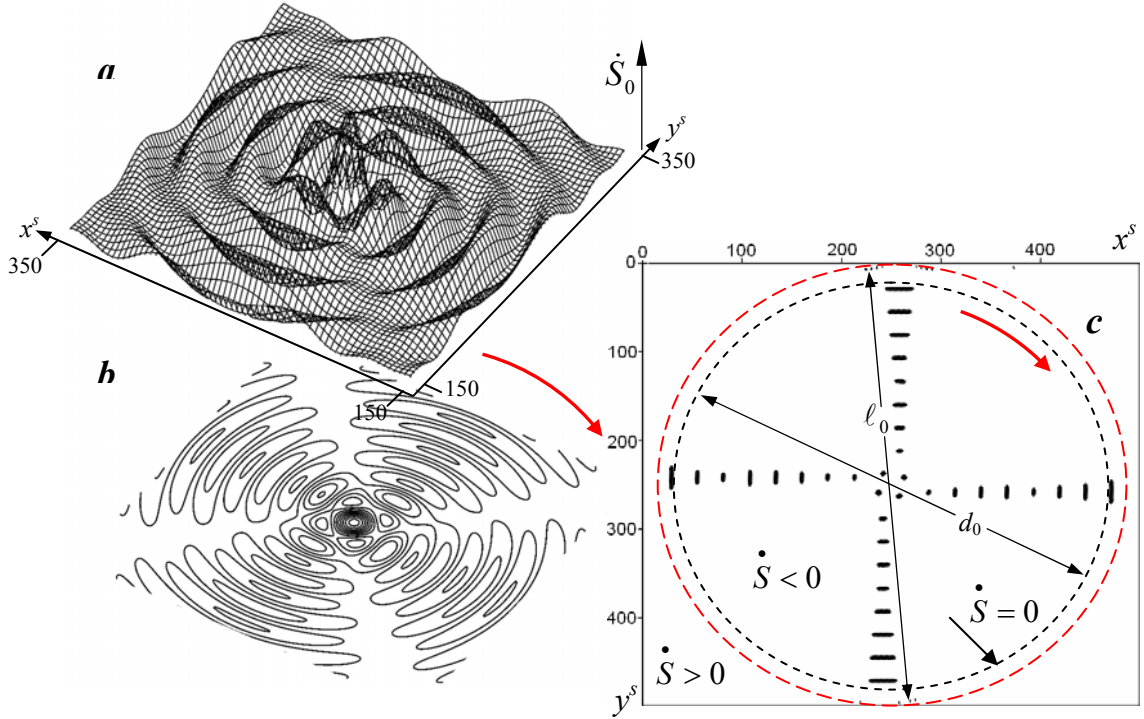

Fig. 16. The total entropy change in the normal to vertical axis flat layer in the tornado steady state as a function of spatial coordinates $x_s$, $y_s$ (*a*) and its isolines projection (*b*) (top view) and self-organization area $dS/dt < 0$ marked with black (*c*). Red arrows indicate the twist direction. Parameters: $h = 1000$ m, where $\rho = 1.087$ $kg/m^3$, $\nu = 1.589*10^{-5}$ $m^2/s$, $\rho^* = 0.887$, $\nu_1^* = 1.182$, $\nu_2^* = 1.182$, $q^* = 1$, $\alpha_1 = 10$, $\alpha_2 = 1$, at the time $\Delta_t = 2.5*10^{-7}$, $\Delta_x = \Delta_y = 0.002$, $t = 2*10^{-4} t_0$. The time step $\Delta_t = 2.5*10^{-7}$; $\Delta_x = \Delta_y = 0.002$. Inner diameter tornado environment, whose estimating is made on the principal of self-organization appearance, $d_0 = 0.884$.

Fig. 16 shows the behavior of the entropy change rate in the plane of the air layer at the altitude of 1000 *m*. The self-organization zone is observed in the basin of the tornado (Fig. 16*c*). The figure shows that there is a boundary region whose diameter $d_0$, where the self-organization thermodynamic conditions (*dS/dt* <0) are no longer hold, resulting in the collapse of the vortices formed outside of the tornado basin arises. At the tornado core (four black points in the center of the Fig. 11c) the entropy change rate takes the highest negative values, which confirms the formation of the most stable and strong vortices near the tornado core. Thus thermodynamic approach confirms the results of transport phenomena modeling and facts observable.

Inner diameter of tornado basin, whose estimating is made on self-organization area, equal to $d_0 \approx 0.884 \ell_0$ which coincides with the value obtained for nonpotential flow from velocity magnitude (5.1.1 Table. 1). The line $\dot{S} = 0$, which shows the outer tornado boundary, outside this line self-organization is absent. There is the formation of dissipative structures – the spiral waves and tornado core – inside.

### 6.4. The free energy change rate

Fig. 17 shows the total free energy change rate

$$\frac{dF^*}{dt} = -\left(\frac{d_e S^*}{dt} + \frac{d_i S^*}{dt}\right),$$

which was calculated according to expression (23). Calculation and figure confirm the of self-organization presence in layers.

In terms of free energy we can also claim two basic self-organization criteria arising from (22): if $\sigma^{e*} < 0$ and $|\sigma^{e*}| \geq \sigma^{i*}$ than the rate is $\dot{F} > 0$. During vortex formation the free energy increases and is transmitted from large to small vortices, in the smallest vortices it dissipates – is transformed into heat.

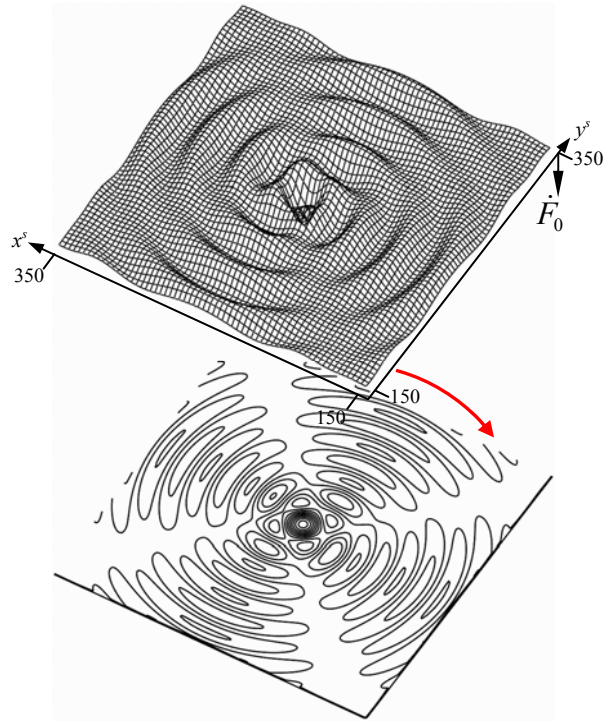

Fig. 17. The free energy change rate in the transverse layer in the steady tornado state as a function of spatial coordinates $x_s$, $y_s$; $h = 1000$ *m*, where $\rho = 1.087$ *kg/m³*, $\nu = 1.589*10^{-5}$ *m²/s*, $\rho^* = 0.887$, $\nu_1^* = 1.182$, $\nu_2^* = 1.182$, $q^* = 1$, $\alpha_1 = 10$, $\alpha_2 = 1$, at the time $\Delta_t = 2.5*10^{-7}$, $\Delta_x = \Delta_y = 0.002$, $t = 2*10^{-4} t_0$. The time step $\Delta_t = 2.5*10^{-7}$; $\Delta_x = \Delta_y = 0.002$.

## 6.5. Tornado vertical velocity component estimation

In evaluating the vertical velocity component $\vartheta_z^* = \vartheta_z / \vartheta_c$ will use the following working hypothesis related to the energy conservation law: The free energy increment in a tornado, which results of the dissipative structures formation (areas of self-organization) in the layer, spends to increase the kinetic energy of vertical air motion

$$F^* - F_0^* = \frac{\rho^* \vartheta_z^{*2}}{2}.$$

We assume that the initial vertical velocity is zero in self-organization area. Knowing the value of the free energy time derivative for self-organization area, where $\dot{F} > 0$ holds, one can express the difference between the values on the left side as follows:

$$F^* - F_0^* = \int_0^{t'} \frac{dF^*}{dt^*} dt^* = \frac{dF^*}{dt^*} \cdot t',$$

here $t'$ – time during which the vertical velocity increments.

Then, from these relations, we obtain the expression for the vertical velocity component:

$$\vartheta_z^* = \sqrt{\frac{2}{\rho^*} \cdot \frac{dF^*}{dt^*} \cdot t'}.$$

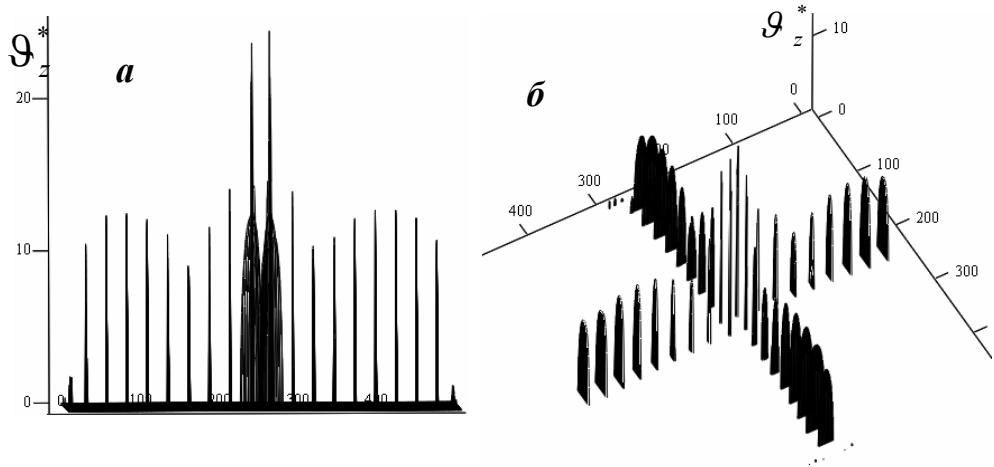

Fig. 18. Estimation of the vertical velocity component in the self-organization area for the time $t' = 0.1$.

Fig. 18 shows the vertical velocity component in the self-organization area for the time $t' = 0.1$. From the figures it is clear that relative values of the vertical component is very high though they are observed only in areas in which the total entropy change is negative. Particularly velocity in the core pipe is significant.

It should be noted that the hypothesis is valid only for the self-organization areas. A more accurate calculation should also take into account the free energy dissipation due to viscous forces and chaotic vortex formation, which is initiated by this velocity. It should also be borne in mind that with increasing layer altitude viscosity increases and the relative velocity values decrease. Therefore, the estimated value of the vertical velocity should be considered maximum.

## 6.6. Characterization of localized structures at height of 10 km. Self-organization changing with the height increases

There is some difficulty to describe the structures at various altitudes. We need to set the initial conditions at these altitudes, but they are unknown. Nevertheless, it was possible to estimate the ratio between the reversible external entropy flows at the altitude of 1 *km* and at the altitude of 10 *km* ($\sigma_0^{e*} = -2000$, Fig. 14; $\sigma_0^{e*} = -10$, Fig. 19).

At the altitude of 10 *km* reversible flow is 200 times less than it at the height of 1 *km*. Exactly last fact indicates that the force of in the steady tornado state is maximal at low altitudes.

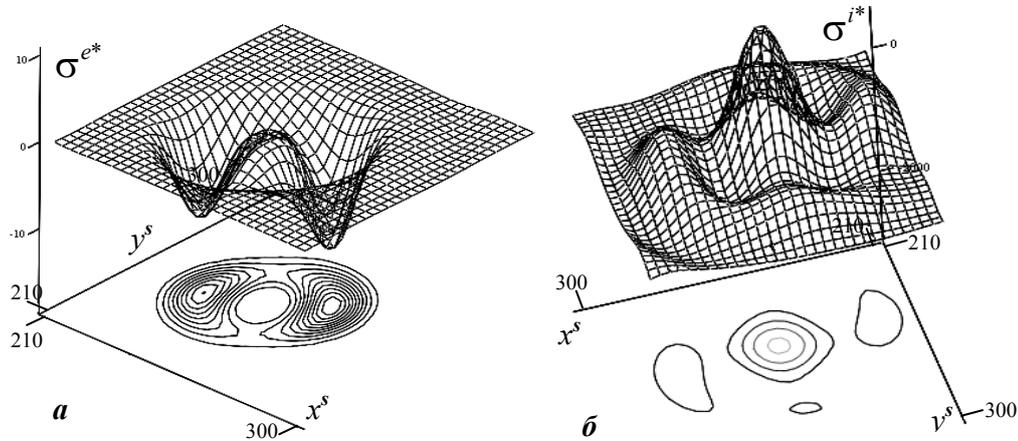

Fig. 19. The functions of external entropy flow (a) and entropy production (b) in the steady tornado state in the arm neighborhood at altitude $h = 10,000$ *m*. Two self-organized mesovortices which twisted around the tornado core and the center dissipation region are clearly visible in (b). This suggests that this altitude is not a limit.

Entropy production has the same tendency – up to 1 *km* ($\sigma_0^{i*} = 4 \cdot 10^4$, in Fig. 15) it is 20 times greater than at 10 *km* ($\sigma_0^{i*} = 2000$, in Fig. 19b). This confirms conclusion above that the basic self-organization process goes more in the lower layers in the steady state. Recall that we are not talking about the tornado initial stage, but about the steady nonequilibrium stage.

At the initial stage vortex arises in the upper atmosphere layers and descends to the bottom. In this case the entropy flows in the arms and in the core of tornado increase (Fig. 14) along with the dissipation growth in the core center (Fig. 15). Going down to the lower atmosphere layers the tornado gain momentum. The merging of vortices swirling around the core occurs to form one large funnel (Fig. 16). With the height increasing the vortices power decreases (Fig. 21,22) and there should be a high point, above which the vortex generating does not occur. This suggests that this energy dissipation growth leads to the tornado collapse. As the tornado develops atmospheric vortices energy decreases due to dissipation of free energy, and when it dries out – the tornado disappears.

Tornadoes may consist of two or more separate arms around the main central twister. Such tornadoes are called *compound tornadoes* and can be almost any power, but most often they are very powerful. They produce extensive damage at vast areas. The derived model captures these different arms (see Fig. 20, Fig. 21). On the Internet you can find pictures of two-arm vortices, the formation of which is explained here.

In such steady state tornadoes, as seen from the figures, supercells with turbulent vortices arise. In the literature it is believed that during the formation of a huge tornado a part of huge thundercloud energy is concentrated in the air volume with the diameter no more than a few hundred cubic meters [15]. Based on the numerical calculation data it is estimated that the energy of an ordinary tornado radius of $l_0/2 = 1$ *km* and an average speed of $\vartheta_c = 46$ *m/s* is comparable with the energy of standard nuclear bombardment $F = 0.96 \cdot 10^{14}$ *J*. (see Table 5 Appendix 2). Single-phase explosion of a nuclear bomb 23 KT power (Trinity polygon, Nevada, 1953) was $F \approx 10^{14} J$ (TNT ton – unit of energy equal to $4,184 \times 10^9$ J) [19].

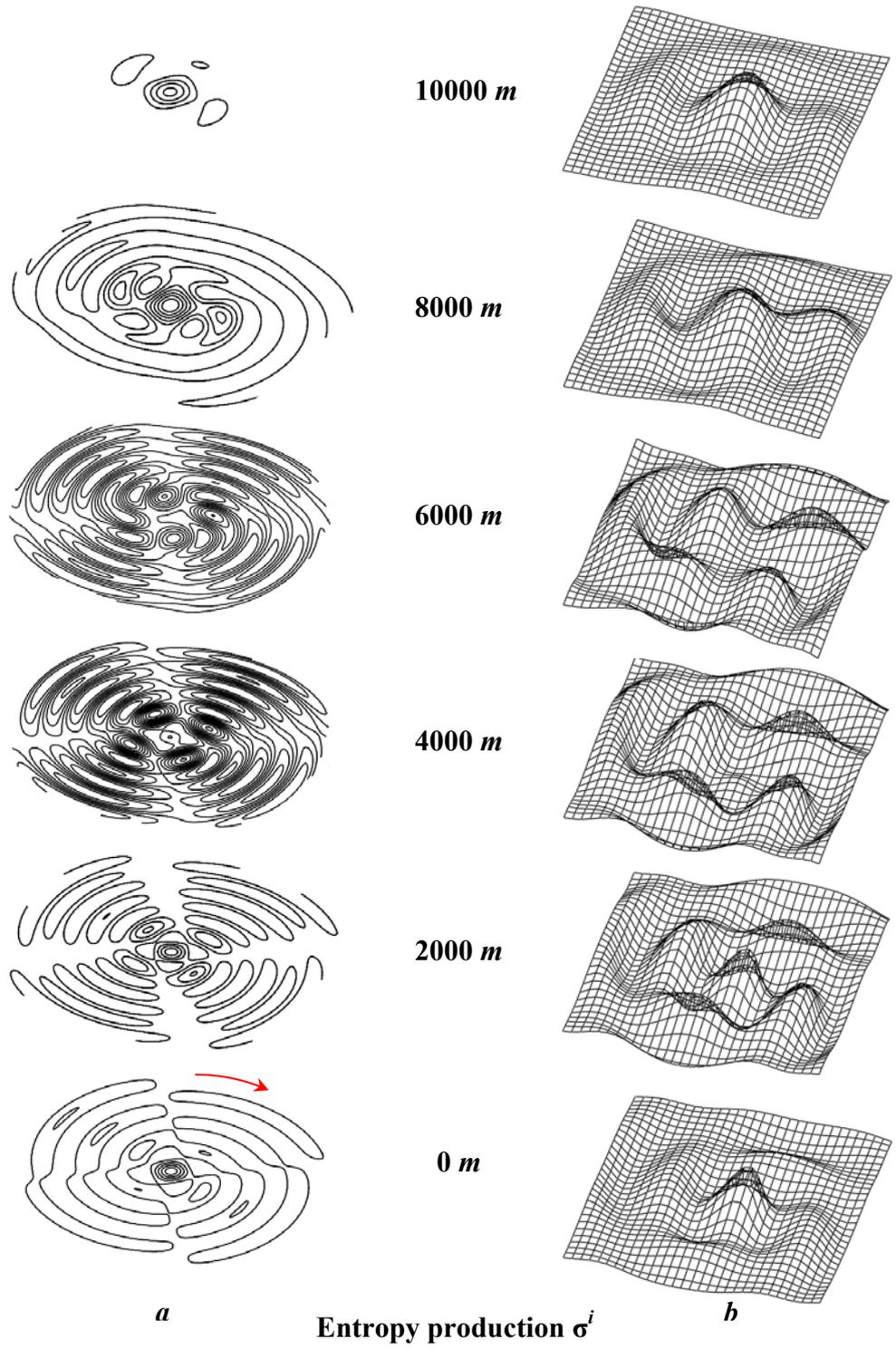

Fig. 20. Entropy production layer by layer from 0 to 10 *km* (*a*) in steady-state tornado mode and increased in 2 times the center region (*b*). Bottom view.

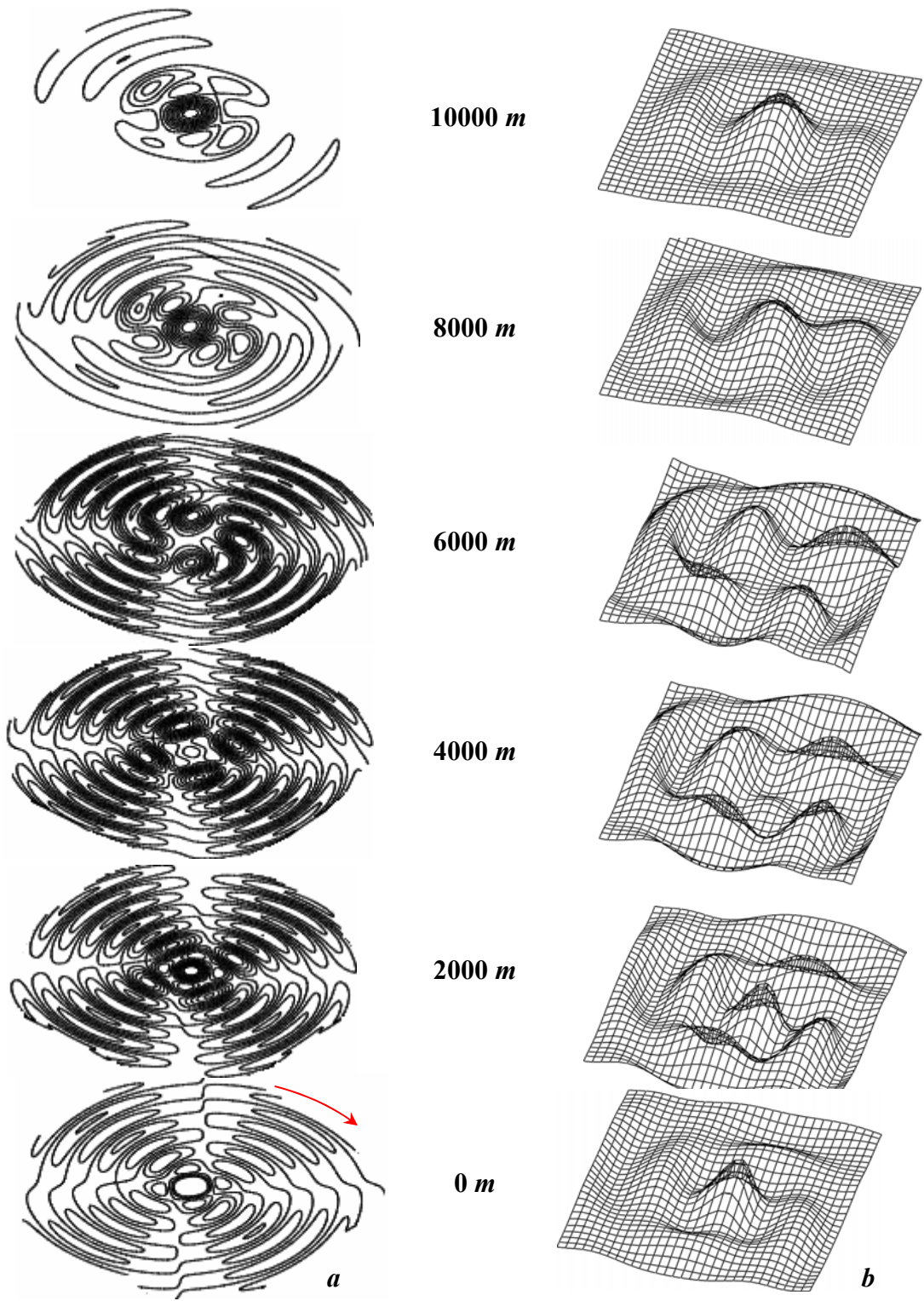

**Entropy change rate** $dS*/dt$

Fig. 21. Layer by layer the entropy change rate in the steady tornado state from 0 to 10 *km* in the equal rate lines form (*a*), and increased by 2 times central region (*b*). Bottom view. At medium altitude (6000 *m*) consisting of two vortices compound tornado are visible.

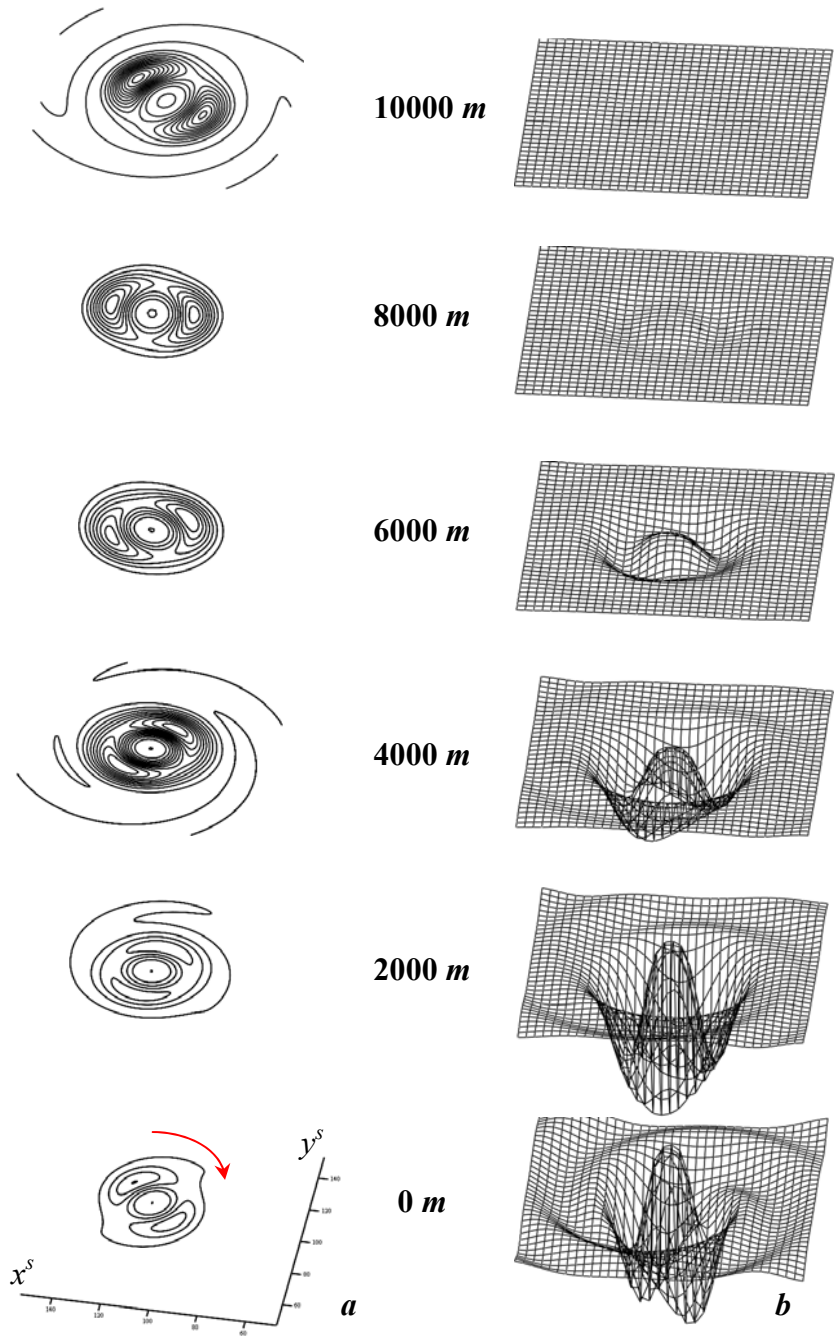

**External entropy flows σ$^{e*}$**

Fig. 22. External entropy flows in the steady tornado state from 0 to 10 *km* layer (*a*) and increased by 1.45 times the central region for each layer(*b*). Bottom view.

Thus the thermodynamic approach with energy conservation law hold allows you to find the conditions of self-organization, identifying not only the minimum rate at which it begins, but also transverse tornado size growth limiting – the entropy change rate boundary condition $\dot{S} = 0$. It does not allow to the tornado transverse dimension grow to infinity. At this boundary competition between the ordering and the collapse of structures leads to the compensation, that ultimately determines the existence of tornado limited size (width). In terms of the energy conservation law we can estimate the limiting values of vertical velocity components, whose values can be calculated exactly in the following research.

## 7. Core formation, depending on the vortices number

By varying the modulus of the topological charge from |m|=1, which corresponds to the central single vortex, up to |m|=5 (5 vortices) is found that the formation of a zone in the increased rotation rate tornado center is occur only at the small number of vortices (1≤|m|≤2). Rapid damping rotation are detected without the formation of one or more core as the number of vortices (|m|>2) in the central part. This suggests that tornadoes with a small number of the central vortex or "tails" are stable, and confirms the fact that only tornado with one or as maximum two "tails" are registered. Formation leading center of concentric radiating circles not considered, and the question of their existence remains open.

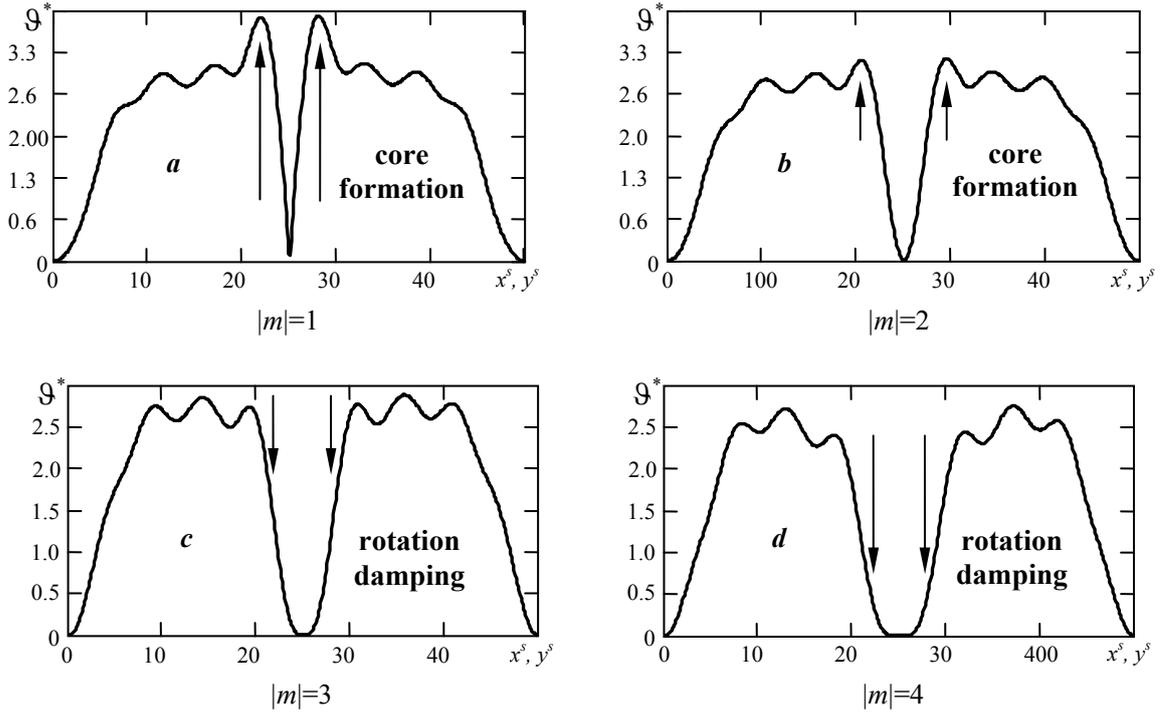

Fig. 23. Diagonal shape of the velocity as a function of the topological charge $m$ module, near the Earth surface ($h=0\ m$) at the time $t=1.9 \cdot 10^{-3}\ t_0$. The core formation depending on the number of tornado vortices is shown. Core forms only with small number of vortices (1−2, (a)−(b)), in other cases the formation of the core does not occur with the tornado central part rotation damping. The initial conditions parameters: $\Phi_0=10$, $R_0=20$.

At Fig. 23 in addition to the tornado core formation other solutions of the Kuramoto-Tsuzuki equation (16), which is not result in a center large rate increasing, are presented. This means that the core in these formations (Fig. 23 c,d) has a specific form, which are not tornado-specific, but cyclone-specific. Any structure absence of the middle zone is observed and this zone can be identified with the cyclone central part (the area where $\vartheta^* < 0.64$ and where the selforganisation does not arise). The latter occurs when the spiral arms number in the screw exceeds 2. All of this means that the cyclones, in center of which self-organization is not observed and the core, like in a tornado, does not arise, also can be described by the Kuramoto-Tsuzuki equation (16). The problem of cyclones describing requires further research. Speeds in these cyclones are modulo much smaller than in the tornado, as shown in Fig. 23.

# 8. Further improvement of the theory

Solution of problems associated with further improvement of the theory not addressed in this article.

1. One should take into account different values of air density in the core (where $\vartheta_z < 0$) and outside (where $\vartheta_z > 0$).

2. The law of energy and mass conservation must be formulated in integral form, with taking into account the peripheral air lowering, which will allow to determine the vertical velocity component $\vartheta_z$.

3. The tornado motion is very unpredictably: although its upper part moves quite predictably with verifiable speed, vortex column lower point "wanders" on the earth surface, sweeping away everything in its way, which fall in the accessibility range. Splinters, soil and debris accumulate in the column bottom, making it more darkly and widely, and this phenomenon is accompanied by objects scattering in all directions with high speed. The mechanism of this rotation was described in this article. However, the creation of such motion video pattern with taking into account layering characteristics calculation is problematic, but it is possible with appropriate computer facilities. Particularly the initial stage tornado formation is interesting, that is outlet to the asymptotic limit.

4. In the transition from the relative motion characteristics to the real characteristics calculation real temperature changes of the $T_h$ – layer average homothermal temperature at height $h$, and also possibly power sources $q_h$ and sinks $\hat{\alpha}_h$ characteristics variation with height should also be considered.

5. If the stress relaxation time $\tau$ – in free molecular regime, which namely takes place at a height, is meaningful, then we should move to the hyperbolic equation of motion (4) and the hyperbolic Kuramoto-Tsuzuki equation.

6. We should take into account the difference between wind speed and its direction at different heights. Tornadoes occur when warm moist air collides with a cold stream of air masses moving at different speed and in the other direction. It is necessary to calculate the spin speed on the initial tornado stage.

7. The strong entropy increasing and the significant vertical velocity component are obtained at the tornado core. At the sufficiently large distance from the tornado tube velocity is directed downwards and external movement contour is closed. The latter means that the internal core flow moves at a slower rate than immediately outside the core, thus in the highest tornado point updraft splits – one air part enters the core, the other moves at the periphery. However, the latter requires rigorous evidence with taking into account of surfaces and also flows descent and lifting speeds.

8. Recent observations show a temporary change of variable electric and magnetic fields created by tornadoes and dust devils [12,13]. It is assumed that it will help to create devices for early tornado warning in the future and also to clarify the geometry and dynamics of vortices. Electromagnetic frequencies are not only generate infrasound, but also modulate the charge density and velocity field and thereby lead to electric and magnetic field fluctuations in the frequency range of 0.5 to 20 Hz, which can be controlled at a distance of several kilometers. Whereas the hydrodynamic tornado characteristics are defined, we can now go to the description of variable electric and magnetic fields.

## References:

1. Turing A. The Chemical basis of morphogenesis // Phyl. Trans. Roy. Soc. L. Ser. B. 1985. Vol. 237. P. 37-72.
2. Arsen'yev S.A., Babkin V.A., Gubar' A.Y., Nikolaevskiy V.N., 2010. The Theory of Meso-Scale Turbulence. Eddies of Atmosphere and Ocean. Publishing House "Regular and Chaotic Dynamics", Moscow -Izhevsk, p. 362 (in Russian with English abstract).

3. Akhromeeva T.S., Kurdyumov S.P., Malinetskii G.G., Samarskii A.A. Nonstationary structure and diffusion chaos .- Moscow: Science. Main Ed. Sci. Lit., 1992. P. 544.
4. Samarskii A.A. Mathematical modeling and computer experiment // Vestn. USSR, 1979, N.5. P. 38-49.
5. Samarskii A.A., Kurdyumov S., Akhromeeva T.S., Malinetskii G.G. Modeling of nonlinear phenomena in modern science // Computer Science and Technology progress.-M.: Science 1987 p.69-91
6. Kuramoto Y. and Tsuzuki T., On the formation of dissipative structures in reaction-diffusion systems, Progr. Theor. Phys., 54, P. 6870-699 (1975).
7. Nicolis G., Prigogine I. Self-organization in nonequilibrium systems. Springer-Verlag, 1973. – P.511
8. Nicolis G., Prigogine I. Exploring Complexity. Springer-Verlag, 1990. -342 Sec.
9. Khromov S.P., Petrosyants M.A. Small-scale vortices. Meteorologyand climatology. Retrieved June 8, 2009.
10. Kikoin I.K. Tables of physical value. Directory. McGraw, 1976, P.1008.
11. "Aviation: Encyclopedia". Moscow: Great Russian Encyclopedia / TsAGI. Zhukovsky, 1994. P.736.
12. Schmitter E.D. Modeling tornado dynamics and the generation of infrasound, electric and magnetic fields // Natural hazards and earth system sciences,2010, 10. p. 295−298.
13. Arsen'yev S.A. Mathematical modeling of tornadoes and squall storms // GEOSCIENCE FRONTIERS, 2011, № 2(2). p.215−221.
14. Zhuravlev V.A. Thermodynamics of irreversible processes in the problems and solutions. Moscow: Nauka, 1979. – P.136
15. http://www.astronet.ru/db/msg/1173645/lect2-4.html
16. http://schools.keldysh.ru/sch1216/materials/sun_sys_do/Wind/wind_2.htm
17. http://hainanwel.com/ru/unusual-world/5 ... souri.html
18. Bystrai G.P. hermodynamics of open systems. Textbook.Ekaterinburg: Izd-vo Ural. State University (AAU neck). 2007. –P.120.
19. http://ru.wikipedia.org/wiki/
20. Samarskii A.A. Computers and Nonlinear Phenomena: Computers and modern science / Auto. foreword. Samarskii A.A. − M.: Nauka, 1988 .- P.192.
21. Malinetskii G.G. Chaos. Structure. Computer experiment. Introduction to nonlinear dynamics. Moscow: Editorial URSS, 2000.- P.256.
22. Dotzek N., Griesler J., Brooks H.E. Statistical modeling of tornado intensity distributions // Atmospheric Research 67-68, 2003. p.163-187.

**Appendix 1**
**Table 1**

**Density $\rho$, pressure $p$, the dynamic viscosity $\eta$ and kinematic viscosity $\nu$ of air at different heights $h$ [10].**

| $h$, m | $\rho(h)$, kg/m$^3$ | $p(h)$, $10^5$·Pa | $\eta(h)$, $10^{-5}$·Pa·s | $\nu(h)$, $10^{-5}$·m$^2$/s |
|---|---|---|---|---|
| 112 | 1.209 | 1 | 1.81 | 1.497 |
| 5477 | 0.612 | 0.51 | 1.828 | 2.985 |
| 22040 | 0.033 | 0.27 | 1.817 | 5.499 |
| 31840 | 1.576·10$^{-3}$ | 0.0013 | 1.77 | 1100 |
| 31890 | 1.464·10$^{-3}$ | 0.0013 | 1.73 | 1100 |

**Table 2**

**Interpolation values of the density $\rho$, pressure *p*, temperature *T*, the length free path $\lambda$, dynamic viscosity $\eta$ and kinematic viscosity $\nu$ of air at different heights *h*, and their dimensionless values.**

| h, km | $\rho(h)$, kg/m³ | $\rho^*(h)$ | T,K | $T^*$ | $p(h)$, 10⁵·Pa | $p^*(h)$ | $\lambda$, 10⁻⁸·m | $\lambda^*$ | $\eta(h)$, 10⁻⁵·Pa·s | $\eta^*(h)$ | $\nu(h)$, 10⁻⁵ m²/s | $\nu^*(h)$ |
|---|---|---|---|---|---|---|---|---|---|---|---|---|
| 0 | 1.22 | 1 | 288.15 | 1 | 1.01 | 1 | 5.81 | 1 | 1.65 | 1 | 1.34 | 1 |
| 1 | 1.09 | 0.89 | 281.65 | 0.98 | 0.89 | 0.89 | 6.50 | 1.10 | 1.73 | 1.05 | 1.58 | 1.18 |
| 2 | 0.96 | 0.78 | 275.15 | 0.95 | 0.79 | 0.78 | 7.07 | 1.22 | 1.80 | 1.10 | 1.87 | 1.39 |
| 3 | 0.85 | 0.69 | 268.65 | 0.93 | 0.70 | 0.69 | 7.82 | 1.35 | 1.88 | 1.14 | 2.22 | 1.65 |
| 4 | 0.74 | 0.61 | 262.15 | 0.91 | 0.62 | 0.61 | 8.68 | 1.49 | 1.95 | 1.19 | 2.62 | 1.95 |
| 5 | 0.65 | 0.53 | 255.65 | 0.89 | 0.54 | 0.53 | 9.66 | 1.66 | 2.02 | 1.23 | 3.10 | 2.30 |
| 6 | 0.57 | 0.47 | 249.15 | 0.86 | 0.47 | 0.47 | 10.78 | 1.86 | 2.09 | 1.27 | 3.66 | 2.72 |
| 7 | 0.50 | 0.40 | 242.65 | 0.84 | 0.41 | 0.40 | 12.07 | 2.08 | 2.15 | 1.30 | 4.33 | 3.22 |
| 8 | 0.43 | 0.35 | 236.15 | 0.82 | 0.36 | 0.35 | 13.55 | 2.33 | 2.20 | 1.34 | 5.11 | 3.80 |
| 9 | 0.37 | 0.30 | 229.65 | 0.80 | 0.31 | 0.30 | 15.25 | 2.63 | 2.25 | 1.36 | 6.04 | 4.49 |
| 10 | 0.32 | 0.26 | 223.15 | 0.77 | 0.26 | 0.26 | 17.24 | 2.97 | 2.28 | 1.39 | 7.14 | 5.31 |

**Table 3**

**Measured external observed variables**

| Measured variables | Values |
|---|---|
| h | 0 m |
| $T_0$ | (288.15 K) |
| $P_0$ | (1.01·10⁵ Pa) |
| $\rho_0$ | (1.225 kg/m³) |
| $\vartheta_c$ | (120 m/s) |
| m | (±1) |

**Table 4**

**Table of transition from dimensionless variables to dimensional variables**

| Dimensionless variables | Dimensional variables |
|---|---|
|  | $t_0 = \ell_0 / \vartheta_c$, $\ell_0 = 1000\,m$, $t_0 = 8.3\,c$, $\vartheta_c = 120\,m/s$ |
| $q^* = 1$ | $q = q^*/t_0$, $q = 0.12\,s^{-1}$ |
| $\alpha_1^* = 10, \alpha_2^* = 1$ | $\hat{\alpha} = \hat{\alpha}^*/t_0\vartheta_c^2$, $\alpha_1 = 8.4\cdot 10^{-5}\,s/m^2$, $\alpha_2 = 8.4\cdot 10^{-6}\,s/m^2$ |
| $\sigma^{i*}, \sigma^{e*}$ | $\sigma^i(h) = \sigma^{i*}(h)\dfrac{\rho_0\vartheta_c^2}{T(h)t_0}$, $\sigma^e(h) = \sigma^{e*}(h)\dfrac{\rho_0\vartheta_c^2}{T(h)t_0}$, $\dfrac{\rho_0\vartheta_c^2}{T(h)t_0} = \dfrac{2125}{T(h)}$ J/K·s·m³ |
| $S^*$ | $S(h) = S^*(h)\dfrac{\rho_0\vartheta_c^2}{T(h)}$, $\dfrac{\rho_0\vartheta_c^2}{T(h)} = \dfrac{17637}{T(h)}$ J/K·m³ |
| $F^*$ | $F(h) = F^*(h)\rho_0\vartheta_c^2$, $\rho_0\vartheta_c^2 = 17637$ J/m³ |

Table 5

Score values of tornado free energy

| Characteristic | Calculation characteristic |
|---|---|
| **Large-scale diameter tornado** | $\ell_0 = 2000\ m$ |
| **The unit volume layer thickness** $dh$ | $dh = \dfrac{V}{\pi R^2}$, at $V = 1\ m^3$, $\dfrac{\ell_0}{2} = 1000\ m$ <br> $dh = 3.18 \cdot 10^{-7}\ m$ |
| **The number of layers** $n$ **at height 0-10 km** | $n = \dfrac{h}{dh}$ <br> $n = 3 \cdot 10^9$ |
| **The value of large-scale velocity** $\vartheta_c$ | $\vartheta_c = 120\ m/s$ |
| The layer free energy $dF$ with the calculated average velocity $\overline{\vartheta}^*$ | $dF = \dfrac{\rho \overline{\vartheta}^2}{2}$, $\overline{\vartheta}^* = 5$, $\overline{\vartheta} = 5 \cdot 120 = 600\ m/s$, <br> $\rho = 1.22\ kg/m^3$ (for layer in height $h=0\ m$), <br> $dF = 220\ kJ$ |
| The column free energy $F$ for height $h$ | $F = n \cdot dF$ <br> $F = 6.6 \cdot 10^{14}\ J$ |
| **The value of large-scale velocity** $\vartheta_c$ | $\vartheta_c = 80\ m/s$ |
| The free energy $dF$ for a layer with the calculated average velocity $\overline{\vartheta}^*$ | $dF = \dfrac{\rho \overline{\vartheta}^2}{2}$, $\overline{\vartheta}^* = 5$, $\overline{\vartheta} = 5 \cdot 80 = 400\ m/s$ <br> $dF = 98\ kJ$ |
| The column free energy $F$ for height $h$ | $F = n \cdot dF$ <br> $F = 2.94 \cdot 10^{14}\ J$ |
| **The value of large-scale velocity** $\vartheta_c$ | $\vartheta_c = 46\ m/s$ |
| The free energy $dF$ for a layer with the calculated average velocity | $dF = \dfrac{\rho \overline{\vartheta}^2}{2}$, $\overline{\vartheta}^* = 5$, $\overline{\vartheta} = 5 \cdot 46 = 228\ m/s$ <br> $dF = 32\ kJ$ |
| The column free energy $F$ for height $h$ | $F = n \cdot dF$ <br> $F = 0.96 \cdot 10^{14}\ J$ |